\documentclass[11pt,a4paper]{article}

\usepackage{microtype}
\usepackage{amsmath, amsfonts,amssymb,amsthm,mathtools,nicefrac,nccmath,cases,physics}
\usepackage{mathrsfs}  
\usepackage{bm}
\usepackage{soul}
\usepackage{dsfont}
\usepackage{booktabs,multirow}
\usepackage{tabularx}
\usepackage[usenames,dvipsnames,table]{xcolor}
\usepackage{colortbl}
\usepackage{caption}
\usepackage{subcaption}
\usepackage{float}
\usepackage{wrapfig}
\usepackage{hhline}
\usepackage{cite}
\numberwithin{equation}{section}
\numberwithin{table}{section}
\usepackage[linktocpage=true]{hyperref}
\hypersetup{colorlinks=true,linkcolor=colorloc3,citecolor=colorloc3,urlcolor=colorloc3}
\usepackage{afterpage}

\def\hybrid{\topmargin -20pt    \oddsidemargin 0pt
	\headheight 0pt \headsep 0pt
	\textwidth 6.5in        
	\textheight 9in         
	\textwidth 6.25in       
	\textheight 9 in       
	\marginparwidth .875in
	\parskip 5pt plus 1pt 
	\jot = 1.5ex
}
\hybrid

\definecolor{colorloc1}{RGB}{164,42,46} 
\definecolor{colorloc2}{RGB}{100,100,100} 
\definecolor{colorloc3}{RGB}{204,119,34}  
\definecolor{colorloc4}{RGB}{25,25,112}  
\definecolor{colorloc5}{RGB}{100,0,0}  
\definecolor{colorloc6}{RGB}{200,200,200}  
\definecolor{colorloc7}{RGB}{70,70,70}  
\definecolor{colorloc8}{RGB}{0,128,128}  

\usepackage[framemethod=default]{mdframed}

\newmdenv[skipabove=10pt,
skipbelow=7pt,
rightline=false,
leftline=true,
topline=false,
bottomline=false,
linecolor=colorloc4,
backgroundcolor=colorloc8!5,
innerleftmargin=4pt,
innerrightmargin=0pt,
innertopmargin=0pt,
leftmargin=2pt,
rightmargin=0pt,
linewidth=2pt,
innerbottommargin=4pt,
frametitlebackgroundcolor=colorloc3]{lbBox}

\newmdenv[skipabove=10pt,
skipbelow=7pt,
rightline=false,
leftline=true,
topline=false,
bottomline=false,
linecolor=colorloc5,
backgroundcolor=colorloc1!5,
innerleftmargin=4pt,
innerrightmargin=4pt,
innertopmargin=0pt,
leftmargin=2pt,
rightmargin=0pt,
linewidth=2pt,
innerbottommargin=4pt,
frametitlebackgroundcolor=colorloc1]{ldBox}
\newenvironment{claimbox}{\begin{ldBox}\vspace{3 mm}
	} {\vspace{1.5 mm}\end{ldBox}}

\usepackage{fancyhdr}

\usepackage{fancyhdr}
\usepackage{sectsty}
\usepackage{titlesec}

\sectionfont{\color{colorloc5}\Large} 

\subsectionfont{\color{colorloc2}\large} 

\subsubsectionfont{\color{colorloc5}} 

\usepackage[linktocpage=true]{hyperref}
\hypersetup{colorlinks=true,linkcolor=colorloc5,citecolor=colorloc4,urlcolor=colorloc4}

\usepackage[shortlabels]{enumitem}

\newcommand{\specialitem}[3][white]{%
	\item[%
	\colorbox{#2}{\textcolor{#1}{\textsf{\makebox(7,7){#3}}}}%
	]
}

\begin{document}


	\baselineskip=14pt
	\parskip 3pt

	\vspace*{-1.5cm}
	
	\vspace{3cm}
	\begin{center}        

	{\color{colorloc5}\bf \huge The Computational Complexity of the \\[0.3em] Weak Gravity Conjecture}   
	\end{center}
	
	\vspace{0.5cm}
	\begin{center}        
		{\bf \large  Stefano Lanza}
	\end{center}
	
	\begin{center}  
		\emph{II. Institut f\"ur Theoretische Physik, Universit\"at Hamburg,\\
			Luruper Chaussee 149, 22607 Hamburg, Germany}
	\end{center}
	
	\vspace{3.5cm}

	\begin{abstract}
		\noindent The Weak Gravity Conjecture imposes stringent constraints on effective field theories to allow for an ultraviolet completion within quantum gravity. While substantial evidence supports the conjecture across broad classes of string theory-derived effective field theories, constructing low-dimensional models realizing it explicitly remains highly non-trivial. In this work, we illustrate how the presence of multiple gauge fields in an effective field theory significantly complicates the bottom-up implementation of the Weak Gravity Conjecture. To this end, we introduce a general algorithm that constructs the convex hull associated with a given set of superextremal states and verifies whether it satisfies the Convex Hull version of the Weak Gravity Conjecture.
		We show that the computational time of this construction grows exponentially with the number of gauge fields, thereby revealing a fundamental obstruction to concrete, algorithmic realizations of the conjecture in theories with many gauge fields.
	\end{abstract}

	\thispagestyle{empty}
	\clearpage
	
	\setcounter{page}{1}


\newpage

\tableofcontents

\newpage


\section{Introduction}
\label{sec:Introduction}

String theory is known to produce an enormous landscape of low-dimensional effective field theories. Certain classes of compactifications are estimated to give rise to approximately $10^{500}$ such effective field theories \cite{Douglas:2003qc,Ashok:2003gk,Douglas:2004zg}, with F-theory compactifications pushing this number as high as $10^{272,000}$ \cite{Taylor:2015xtz}. 
The sheer magnitude of this landscape renders any systematic, direct exploration aimed at extracting phenomenological insights practically unfeasible.

In the recent years, the Swampland program has offered an alternative, antithetical route to try to extract relevant phenomenological information out of the large landscape of string theory effective field theories, without passing through direct scans thereof.
Initiated with the seminal works \cite{Vafa:2005ui,Arkani-Hamed:2006emk,Ooguri:2006in}, the program pursues a \emph{bottom-up} approach.
The idea is to first identify universal features of consistent effective theories derived from string theory; then, these are elevated to conjectural criteria that any low-dimensional effective field theory that originates from string theory ought to obey.
Thus, equipped with these criteria -- the \emph{Swampland conjectures} -- one could devise low-dimensional effective field theories with desired phenomenological properties, and that are allegedly consistent with an ultraviolet completion within string theory.

However, the Swampland conjectures are speculative in nature. 
The vastness of the landscape precludes comprehensive algorithm scans to confirm their validity, and their generality makes them difficult to establish rigorously.
Nevertheless, the Swampland program has yielded substantial insights into both the phenomenological and geometric structure of broad classes of string theory-derived effective theories, as for instance reviewed in \cite{Palti:2019pca,Agmon:2022thq}.

At the same time, the Swampland conjectures impose severe, sophisticated constraints on the low-energy effective field theories, which are oftentimes hard to check.
This raises an important question: are these conjectures computationally more tractable than the direct scanning of the landscape they aim to circumvent?
That is, if the conjectures are to be useful in practice, one must ask: \emph{How computationally feasible is it to verify whether a given effective theory satisfies a Swampland conjecture?} And relatedly, \emph{Is it easier to construct a consistent effective theory obeying the Swampland constraints, or to search directly for a theory in the landscape with desired properties?}

In this work, we address these questions by focusing on one of the cornerstones of the Swampland program, the Weak Gravity Conjecture.
Originally proposed in  \cite{Arkani-Hamed:2006emk}, the conjecture posits that any effective field theory consistent with quantum gravity must include superextremal particles -- particles whose charge-to-mass ratios exceed that of extremal black holes.
This is indeed necessary to enforce the decay of extremal black holes, consequently avoiding the remnant problem, while being consistent with the absence of global symmetries.
Since the original formulation, the conjecture has undergone several refinements. 
Notably, in \cite{Palti:2017elp} the conjecture has been extended so as to make manifest the role of interacting scalar fields, while in \cite{Cheung:2014vva} it has been illustrated how the original conjecture should be extended whenever several gauge fields dynamically appear in the low-energy effective field theory, a refinement named the \emph{Convex Hull Weak Gravity Conjecture}. 
The latter asserts that, in a given effective field theory, the black hole extremality region ought to be fully covered by an appropriately defined convex hull generated by a set of superextremal particles, a stronger requirement than the one portrayed by the original Weak Gravity Conjecture.
We refer to the reviews \cite{Palti:2020mwc,Harlow:2022ich} for further details, and developments revolving around the conjecture.

Here, we will not delve into the motivations supporting the Weak Gravity Conjecture, or try to prove any of its inherent statements; rather, we will \emph{assume} the validity of the Weak Gravity Conjecture -- in particular in its Convex Hull formulation.
Instead, we will focus on \emph{algorithmic} aspects of the conjecture, and the core question that we will address is: \emph{How hard is it to determine whether a theory with multiple gauge fields obeys the Convex Hull Weak Gravity Conjecture employing an algorithm?}

This question lies at the interface of theoretical physics and computability theory. 
Computability theory, at intersection of logic and computer science, offers the right tools to inquire about the computational feasibility of a given problem by a machine.
At the center of computability theory is determining the problems that are computable by a \emph{Turing machine}, a simple, idealized machine that performs operations in sequence by writing symbols on an infinitely long tape. 
Here, we shall be interested in the computability as measured by the \emph{time complexity}: this consists in the number of steps that an algorithm needs to halt, and output the answer, as a function of the size of the input.
The problems that are considered tractable by a Turing machine are those for which the time complexity is a \emph{polynomial} function in the input size. 
By contrast, there exist several, important problems that cannot be solved in such a polynomial time: for instance, they may require a number of steps that is \emph{exponential} in the input size -- the \emph{traveling salesperson problem} is a notable example thereof -- or they may be undecidable in the first place -- as the \emph{halting problem}.

In string phenomenology context, computability theory has helped to assess whether the problem of finding vacua is feasible by a Turing machine \cite{Denef:2006ad,Denef:2017cxt,Halverson:2018cio}. 
The strategy employed relies on relating string phenomenological questions to known computational problems, for which the time complexity is known.
Here, as we shall see, the realization of the Convex Hull Weak Gravity Conjecture in presence of multiple gauge fields reduces firstly to a \emph{convex hull construction problem}, and then to a \emph{set containment problem}, which are well-studied questions in computability theory -- see, for instance, \cite{Chazelle1993,BDH,Avis1997,Eaves1982,Freund1985,Gritzamann1994,GKL} for a sample of works addressing these problems.
Building on these earlier works, we can infer whether a machine could find realizations of the conjecture in polynomial time.

To this end, in Section~\ref{sec:WGCr} we present an explicit algorithm designed to test the Convex Hull Weak Gravity Conjecture in a given effective field theory, and find the specific states that concur in its realization.
We then extend this framework to explore \emph{minimal} realizations of the Weak Gravity Conjecture, seeking the smallest possible sets of superextremal states satisfying the conjecture.
The main conclusion that we will draw is the following: the time needed by a Turing machine to construct the convex hull out of a set of superextremal states, and then test the Convex Hull Weak Gravity Conjecture grows \emph{exponentially} with the number of gauge fields, implying that for a large number of gauge fields convex hull constructions, and subsequent tests of the conjecture are computationally infeasible.
More efficient checks of the Convex Hull Weak Gravity Conjecture can be obtained provided that one renounces to construct the convex hull in the first place, or by restricting the number of dynamical gauge fields appearing in the effective theory.

This work is articulated as follows.
In Section~\ref{sec:WGC} we overview the assertions of the Weak Gravity Conjecture in presence of several gauge fields; we explain why testing the Weak Gravity Conjecture is computationally involved, and we overview the major computational complexity classes.
In Section~\ref{sec:WGCr} we present the \textsf{WGCHull} algorithm tailored to perform checks of the Weak Gravity Conjecture in a given effective field theory, and we give a quantitative estimate of the computational complexity of the problem of testing the Weak Gravity Conjecture.
In Section~\ref{sec:WGCm} we investigate whether extracting minimal realizations of the Weak Gravity Conjecture, involving the least possible number of superextremal particles, can be done in polynomial time.
Finally, in Appendix~\ref{sec:Hull_gen} we provide additional details about the algorithmic construction of the convex hull in presence of several gauge fields.

\section{Why the Weak Gravity Conjecture is complex}
\label{sec:WGC}

Proposed in \cite{Arkani-Hamed:2006emk}, the Weak Gravity Conjecture is one of the founding pillars of the Swampland Program.
In this section, we review the motivation and the statement of the Weak Gravity Conjecture as originally conceived in \cite{Arkani-Hamed:2006emk}, and some subsequent refinements that are crucial for the discussion of the forthcoming sections. 
We further highlight why the Weak Gravity Conjecture is computationally complex, and how to classify the computational complexity of algorithmic problems.

\subsection{The Weak Gravity Conjecture and convex hulls}
\label{sec:WGC_CH}

Consider an effective field theory in generic $D > 3$ spacetime dimension, coupled to gravity and containing a single $U(1)$ gauge field, represented as the vector $A^\mu$.
The action of such an effective field theory contains the terms
\begin{equation}
	\label{eq:WGC_S}
	S_D = \frac{M_{\rm P}^{D-2}}{2} \int {\rm d}^D x \sqrt{-g} \left( R - \frac1{2 g^2 M_{\rm P}^2} F_{\mu\nu} F^{\mu\nu}  + \ldots \right)\,.
\end{equation}
Here $M_{\rm P}$ denotes the $D$-dimensional Planck mass, and $R$ is the Ricci scalar; furthermore, $F_{\mu\nu} = 2 \partial_{[\mu}  A_{\nu]}$ is the field strength of the gauge field $A^\mu$, and $g$ is its dimensionless coupling.  

Additional terms -- including matter or higher-derivative interactions -- may appear in the effective theory, but for the moment we focus exclusively on the couplings between gravity and the $U(1)$ gauge field as captured by the terms explicitly written in \eqref{eq:WGC_S}.
The action \eqref{eq:WGC_S} admits Reissner-Nordstr\"om black hole solutions, characterized by mass $M$ and global $U(1)$ charge $Q$.
In order to avoid naked singularities, the mass $M$ and charge $Q$ of such black hole solutions need to obey the condition
\begin{equation}
	\label{eq:WGC_subextr}
	M_{\rm P} \frac{g |Q|}{M}  \leq \alpha \,,  \quad \text{with} \quad \alpha = \sqrt{\frac{D-3}{D-2}} \,.
\end{equation}
Black hole solutions that obey the strict inequality in \eqref{eq:WGC_subextr} are said to be \emph{subextremal}, and are characterized by two separate horizons; instead, solutions that saturate the bound \eqref{eq:WGC_subextr} are termed \emph{extremal}, and possess a single horizon.

The Weak Gravity Conjecture asserts that any low-energy effective field theory admitting an ultraviolet completion within string theory needs to contain a \emph{superextremal particle}, with charge-to-mass ratio
\begin{equation}
	\label{eq:WGC}
	M_{\rm P} \frac{g |q|}{m} \geq \alpha \,.
\end{equation}
Indeed, the existence of such a particle ensures that the decay of an extremal black hole into a subextremal one via emission of this particle is kinematically allowed.

In string-theoretic contexts, low-energy effective field theories often contain multiple $U(1)$ gauge fields, naturally arising through the dimensional reduction of higher-dimensional effective field theories.
These may include gauge fields that are not currently observed, such as gauge fields belonging to dark sectors.
The generalization of the Weak Gravity Conjecture to effective field theories endowed with multiple gauge fields is however non-trivial, and some (partly equivalent) formulations have been proposed.
In this work, we will focus on the \emph{Convex Hull Weak Gravity Conjecture} introduced in \cite{Cheung:2014vva}.

Consider a $D$-dimensional gravitational effective field theory, endowed with $N$ abelian gauge fields $A^I_\mu$, with $I = 1, \ldots, N$.
Their low-energy, minimal interactions with gravity are described by an action of the form
\begin{equation}
	\label{eq:WGC_S2}
	S_D = \frac{M_{\rm P}^{D-2}}{2} \int {\rm d}^D x \sqrt{-g} \left( R - \frac1{2 M_{\rm P}^2} f_{IJ} F^I_{\mu\nu} F^{J \,\mu\nu}  + \ldots \right)\,.
\end{equation}
Here, $f_{IJ}$ is the gauge kinetic matrix; for simplicity, we shall assume that the gauge kinetic matrix $f_{IJ}$ is diagonal, and normalized such that $f_{IJ} = \delta_{IJ}$.
Additional terms may appear in the action \eqref{eq:WGC_S2}, which would introduce other interactions with either gravity, or with the gauge fields $A^I_\mu$; as before, for the moment, we restrict our attention to the interactions explicitly written in \eqref{eq:WGC_S2}.

The action \eqref{eq:WGC_S2} admits Reissner-Nordstr\"om black hole solutions.
A black hole solution with mass $M$, and elementary electric charges ${\bf Q} = (Q_I)^T$, with $Q_I$ being the elementary electric charge with respect to the gauge field $A^I_\mu$, is devoid of naked singularities provided that
\begin{equation}
	\label{eq:WGC_extr2}
	M_{\rm P} \frac{\| {\bf Q} \| }{M} \leq \alpha
\end{equation}
where $\| {\bf Q} \|^2 = \sum\limits_{I = 1}^N Q_I^2$, and $\alpha$ as in \eqref{eq:WGC_subextr}.
Again, extremal solutions are those that saturate the bound in \eqref{eq:WGC_extr2}, otherwise they are termed subextremal.

The main motivation behind the Weak Gravity Conjecture is the requirement that any charged black hole, and specifically extremal ones, need to decay.
However, as noticed in \cite{Cheung:2014vva}, it is not enough to require that \eqref{eq:WGC} holds for each of the $U(1)$ charges separately, for a more stringent condition needs to be obeyed.
Indeed, let us assume that there exist several species of particles, with each of the species characterized by particles of mass $m^{(i)}$, and elementary electric charges ${\bf q}^{(i)}$, with $i = 1, \ldots, N_s$; we further assume that they are superextremal; namely $M_{\rm P} \frac{\| {\bf q}^{(i)} \| }{m^{i}} \geq \alpha$ for all the species.
Then, let us consider the extreme case where the black hole charge is fully discharged, by emitting $n^{(i)}$ particles of each of the $i$-th species.
In order for this decay channel to be allowed, it is necessary that
\begin{equation}
	\label{eq:WGC_decQM}
	 {\bf Q} = \sum\limits_{i = 1}^{N_s} n^{(i)} {\bf q}^{(i)} \,, \qquad M \geq \sum\limits_{i = 1}^{N_s}  n^{(i)} m^{(i)}\,,
\end{equation}

To make the structure clearer, as in \cite{Cheung:2014vva}, it is convenient to introduce the charge-to-mass ratio vector for the black hole ${\bf Z} = M_{\rm P} \frac{{\bf Q}}{M}$, so that an extremal black hole is characterized by the vector 
\begin{equation}
	\label{eq:WGC_BHextr_b}
	 \|{\bf Z}_{\rm ext}\|  = \alpha \,,
\end{equation}
with $\alpha$ as in \eqref{eq:WGC_subextr}; we notice that the relation \eqref{eq:WGC_BHextr_b} determines a sphere in ${\bf z}$-space, that we name \emph{black hole extremality sphere}.
Analogously, for each of the particle species, we introduce the charge-to-mass ratio vectors ${\bf z}^{(i)} = M_{\rm P} \frac{{\bf q}^{(i)}}{m^{(i)}}$. 
Then, defining $\sigma^{(i)} = \frac{n^{(i)} m^{(i)}}{M}$, the conditions \eqref{eq:WGC_decQM} for allowing the decay become
\begin{equation}
	\label{eq:WGC_decZM}
	{\bf Z} = \sum\limits_{i = 1}^{N_s} \sigma^{(i)} {\bf z}^{(i)} \,, \qquad  \sum\limits_{i = 1}^{N_s} \sigma^{(i)} \leq 1 \,.
\end{equation}
These conditions express the fact that the black hole charge-to-mass ratio vector ${\bf Z}$ has to reside within the \emph{convex hull} generated by the charge-to-mass ratio vectors ${\bf z}^{(i)}$ of the particles species, augmented by the origin.
In fact, let us recall that the convex hull generated by a set of vectors $\mathcal{W} = \{{\bf v}^{(l)}\}_{i = l,\ldots, M}$ is the smallest convex set that contains $\mathcal{W}$, and can be defined starting from its elements as 
\begin{equation}
	\label{eq:WGC_convS_0}
	{\rm conv} (\mathcal{W}) := \left\{ {\bf z} \in \mathbb{R}^N : \; \exists\, \lambda^{(l)} \in [0,1] \; \text{such that}\;  \sum\limits_{l = 1}^{M} \lambda^{(l)} = 1 \; \text{and}\; {\bf z} = \sum\limits_{l = 1}^{M} \lambda^{(l)} {\bf z}^{(l)}   \right\} \,.
\end{equation}
Notice that, in our case, since the black hole subextremality region $\| {\bf Z} \| \leq \alpha$ trivially contains the origin ${\bf z}_{\rm o}$ of the ${\bf z}$-space, we shall be interested in convex hulls that include ${\bf z}_{\rm o}$ in the first place; moreover, the convex hull ought to be also generated by the set of superextremal particles $\mathcal{S} = \{{\bf z}^{(i)}\}_{i = l,\ldots, N_s}$.
Thus, choosing ${\bf v}^{(l)} =\{ {\bf z}^{(i)}, {\bf z}_{\rm o}\}$, we get
\begin{equation}
	\label{eq:WGC_convS}
	{\rm conv} (\mathcal{S} \cup {\bf z}_{\rm o}) = \left\{ {\bf z} \in \mathbb{R}^N : \; \exists\, \sigma^{(i)} \in [0,1] \; \text{such that}\;  \sum\limits_{i = 1}^{N_s} \sigma^{(i)} \leq 1 \; \text{and}\; {\bf z} = \sum\limits_{i = 1}^{N_s} \sigma^{(i)} {\bf z}^{(i)}   \right\}\,,
\end{equation}
where we have further relabeled $\lambda^{(i)} = \sigma^{(i)}$, with $\sigma^{(i)}$ as in \eqref{eq:WGC_decZM}.

\begin{figure}[t]
	\centering
	\includegraphics[width=0.7\textwidth]{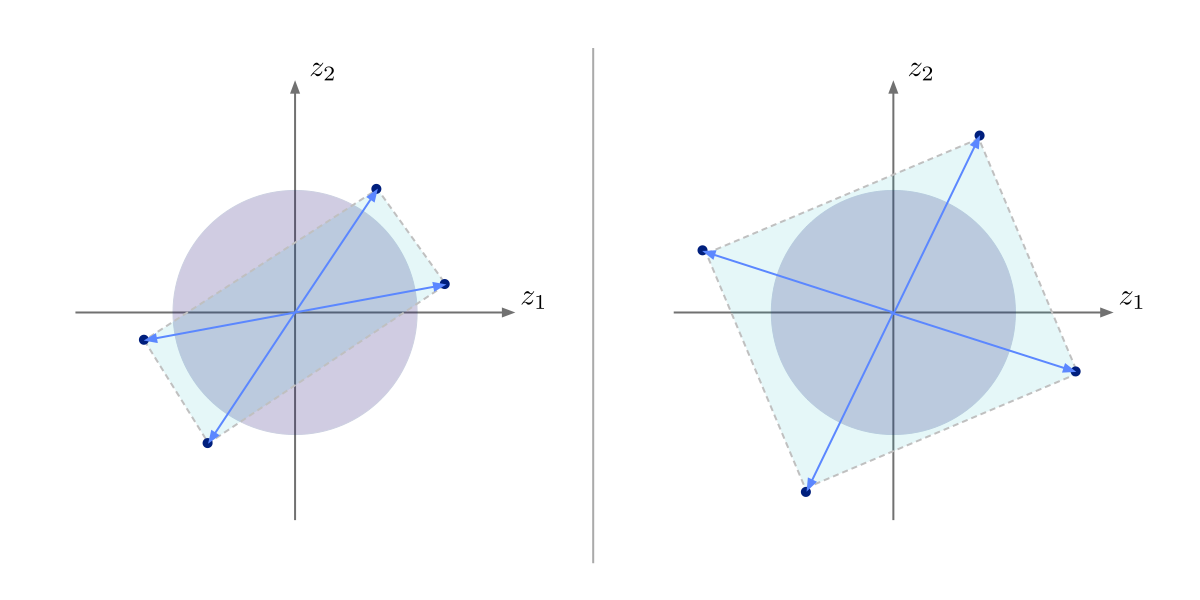} 
	\caption{\footnotesize On the left, an example of four particles not realizing the Convex Hull Weak Gravity Conjecture: the four particles (blue dots) in the $(z_1, z_2)$ plane, generate a convex hull (the light blue rectangle), that does not fully cover the subextremality disk, depicted as the blue disk. On the right, an example of four particles that determine a convex hull that fully covers the subextremality disk, thus realizing the Convex Hull Weak Gravity Conjecture.		
		\label{Fig:WGC_CH}}
\end{figure}

Hence, the Convex Hull Weak Gravity Conjecture asserts that the entire subextremal region (a ball of radius $\alpha$ in $\mathbf{Z}$-space) must lie within the convex hull \eqref{eq:WGC_convS}.
Such a requirement is indeed stronger than requiring \eqref{eq:WGC} along some charge directions.
For instance, in Figure~\ref{Fig:WGC_CH}, on the left, is depicted a case where, despite the four particles are superextremal, they cannot guarantee the decay, for there are some black holes with charge-to-mass ratio vectors ${\bf Z} \notin {\rm conv} (\mathcal{S})$. 
On the right, is a depiction of four superextremal particles fully covering the black hole subextremality region, as any ${\bf Z} \in {\rm conv} (\mathcal{S})$.

The representation of a convex hull via its generating vectors that we have just presented is known as the $\mathcal{V}$-representation. 
Alternatively, one may describe the same hull, or rather its defining polytope using the $\mathcal{H}$-representation, defined as the intersection of a finite number of half-spaces. 
Let $\mathcal{H}_\alpha$ be the hyperplane
\begin{equation}
	\label{eq:WGC_conv_cHa}
	\mathcal{H}_\alpha: \;{\bf a}^T_{\alpha} {\bf z} + 1 = 0 \qquad \text{with $\alpha = 1, \ldots, m$}\,,
\end{equation}
for some vectors ${\bf a}_\alpha$, and let $H_\alpha$ be the corresponding half-space:
\begin{equation}
	\label{eq:WGC_conv_Ha}
	H_\alpha = \left\{ {\bf z} \in \mathbb{R}^N \;|\; {\bf a}^T_{\alpha} {\bf z} + 1 \geq 0 \right\}\,.
\end{equation}
The convex hull of a set of points $\mathcal{W}$ can then be equivalently represented as
\begin{equation}
	\label{eq:WGC_conv_Hrep}
	{\rm conv} (\mathcal{W}) =  \bigcap\limits_{\alpha = 1}^m H_\alpha \,.
\end{equation}

In sum, for multiple gauge fields, the Weak Gravity Conjecture becomes a two-step problem: firstly, it is a \emph{convex hull construction problem}, about determining the convex hull that the set of superextremal particles build; then, it is a \emph{containment problem}, consisting of checking whether a ball (namely, the black hole subextremality region) is enclosed within a polytope (in our case, associated with the convex hull generated by the set of superextremal particles).
Such problems are typically algorithmically involved, especially if the number of gauge fields, which sets the dimension of the charge-to-mass ratio vector space, is very large.
Hence, it could be useful to have at hands an appropriate algorithm so that, given a set of superextremal particles, putative candidates for realizing the Weak Gravity Conjecture, it tells whether their charge-to-mass ratio vectors could appropriately cover the black hole subextremality region, and identifies the particles that realize the hull.

However, assuming to build an algorithm capable of performing checks of the Convex Hull Weak Gravity Conjecture, how efficient could it be?
For instance, do we expect such a tailored algorithm to \emph{halt} after a reasonable amount of time, so as to give an answer about whether the Weak Gravity Conjecture is realized? 
Or would the algorithm inevitably need an exponential amount of time to halt?  

This question falls within the domain of computability theory. Notably, some containment problems of this sort have been shown to be among the most computationally challenging problems \cite{Eaves1982,Freund1985,GKL}. 
By leveraging results from that literature, we can assess the algorithmic feasibility of verifying the Convex Hull Weak Gravity Conjecture, shedding light on the computational complexity inherent in checking quantum gravity consistency conditions.

\subsection{Classifying the complexity of problems}
\label{sec:WGC_compl}

As discussed at the end of the previous section, our goal is to determine whether verifying or refuting instances of effective field theories according to whether they obey the Weak Gravity Conjecture can be done efficiently. But what precisely do we mean by `efficiently', and how can we quantify it?

Computational complexity theory -- at the crossroads of mathematical logic and theoretical computer science -- provides the tools to rigorously address this question. It allows us to characterize which problems are feasibly solvable by algorithms and which lie beyond practical reach. 
For a comprehensive introduction, we refer the reader to standard textbooks such as \cite{arora_barak_2009,Goldreich_2010}. 
In this section, we outline the key concepts and definitions that we will rely on throughout the remainder of this work.

\noindent\textbf{Types of problems.} Any computational task must begin with a well-posed problem. 
Broadly speaking, such problems fall into two categories: \emph{decision problems} and \emph{search problems}. 
Decision problems are boolean problems, which can be answered with a `Yes', or a `No', \emph{tertium non datur}. For instance, ``\textit{Is the integer $N$ prime?}'', or ``\textit{Does the set $\mathcal{S}$ contain the element $a$?}''. 
In contrast, search problems require one to find a solution that satisfies a given criterion, such as ``\textit{What are the roots of this equation?}'', or ``\textit{What is the shortest path between two locations?}''.

\noindent\textbf{Measuring complexity.}  Once a problem is specified, computational complexity theory asks: \emph{how many `resources' are required in order to solve the problem?} 
Indeed, the amount of resources required measures the `\emph{complexity}' of the problem.
The term `\emph{resources}' may refer to different quantities, such as time, space, parallelism;  accordingly, different notions of complexity can be introduced.
In particular, based on tame geometry \cite{van-den-Dries:1988}, within quantum field theories, either gravitational or not,  one can introduce a measure of complexity that is inherent to the theory \cite{Grimm:2023xqy,Grimm:2024elq,Grimm:2025lip}.
Here, however, we focus on \emph{time complexity}.

The natural way to measure such a time complexity is counting the number of steps that the machine requires to halt, and then express this number as a function of the size of the input.
Namely, given an input of size $n$, the time complexity can be quantified via a (positive-definite) function $T(n)$, counting the steps that the machine needs before halting.

Generically, the ways in which modern machines -- for instance, ordinary computers -- operate are highly non-trivial; moreover, it may appear that the diversity in their architectures renders it difficult to define a \emph{general} notion of time complexity.
In this regard, the \emph{Church-Turing thesis}, widely believed to be correct \cite{arora_barak_2009,Goldreich_2010}, comes to help, proposing a notion of universality in detecting the problems that can be efficiently addressed by a machine.
The thesis states that any function can be computed by any machine if and only if it can be computed by a Turing machine.
Loosely speaking, a \emph{Turing machine} is a simple mathematical model of computation that manipulates symbols on a tape according to a finite set of rules. 
By virtue of this universality, it suffices to analyze problems using Turing machines, and results obtained in this setting are robust across different architectures.

It is worth stressing that such a definition of time complexity refers to a \emph{problem}, and not to a specific algorithm that ought to computationally solve that problem.
Indeed, ideally, the computational complexity of a problem is measured by the most efficient algorithm that solves it.
In practice, however, even upper or lower bounds obtained via specific algorithms are informative.

\begin{figure}[t]
	\centering
	\includegraphics[width=0.35\textwidth]{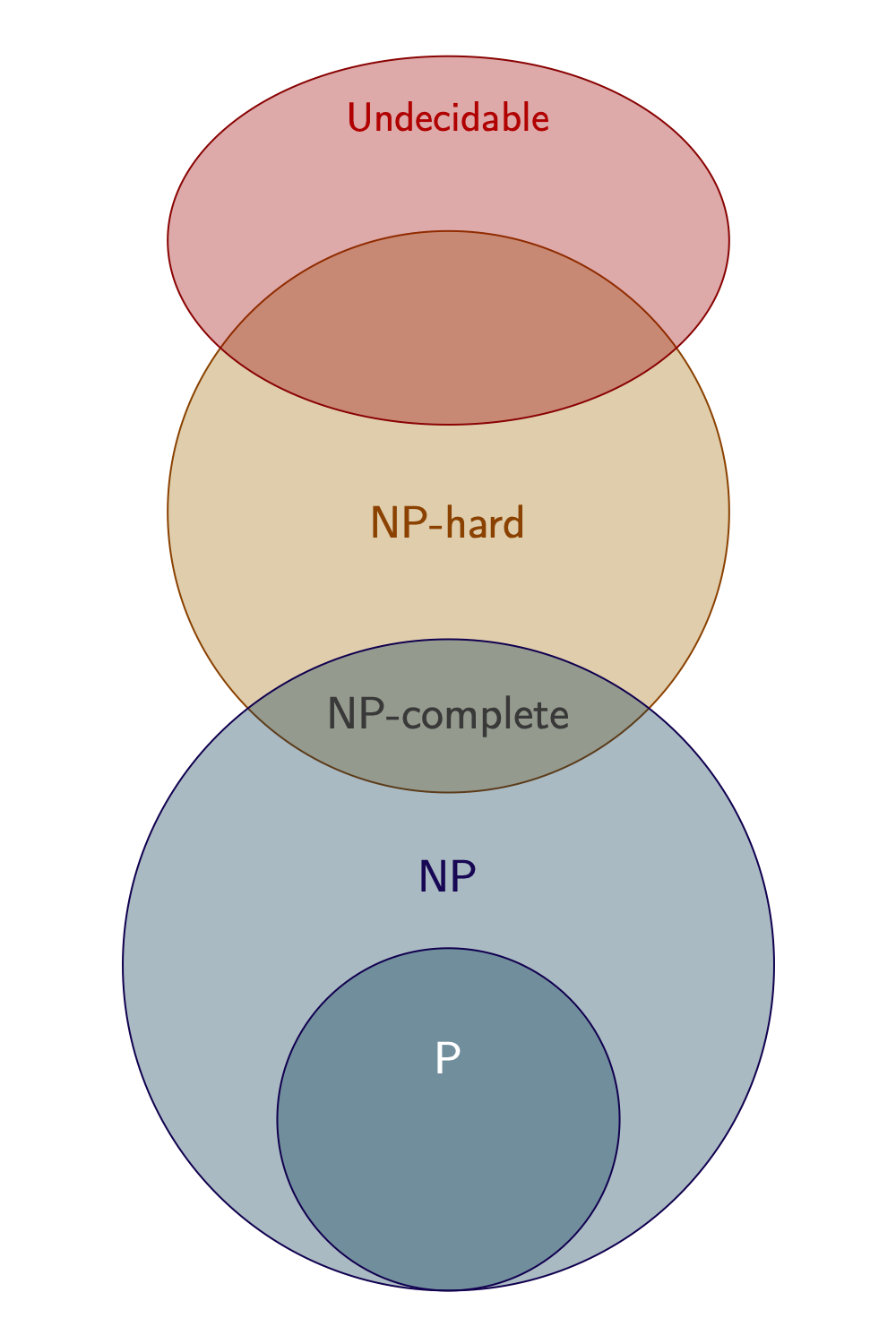}
	\caption{\footnotesize Intersection of basic computational complexity classes.
		\label{Fig:Compl_Classes}} 
\end{figure}

\noindent\textbf{Complexity classes.} Problems can be grouped into \emph{complexity classes}, defined by the asymptotic behavior of their time complexity function $T(n)$ in terms of the input size $n$. 
The following classes are central to our work (see \cite{arora_barak_2009} for a more detailed discussion):
\begin{description}[wide, labelwidth=!, labelindent=0pt, itemsep=0mm]
	\item[\textsf{P}, or polynomial time:] Problems in this class can be solved in time bounded by a polynomial function of the input size: $T(n) = \mathcal{O}(n^c)$ for some $c \geq 0$.  
	Examples of questions belonging to this class include very simple problems, such as computing the sum of two integers, or more complicated ones, such as determining whether a number is prime.
	Problems in this class are generally regarded as \emph{tractable}, or realistically solvable by machines.
	\item[\textsf{EXP}, or exponential time:] This class contains all the problems that a Turing machine requires an exponential time to solve; namely, the number of steps before halting behaves as $T(n) \sim \mathcal{O}(2^{n^c})$, for some $c > 0$.
	The problems that belong to this class are instead considered to be \emph{intractable} by a machine, as their runtime grows too quickly to be feasible for even moderately large inputs.
	\item[\textsf{NP}, or non-deterministic polynomial time:] problems belonging to this class are those for which, \emph{given a solution}, its correctness can be verified in polynomial time.
	For instance, for search problems, given a known solution, there exists an algorithm that checks that it is an actual solution in polynomial time.
	Or, for decision problems, we can think of them as determining whether a given instance belongs to the `Yes', or `No' subset. 
	A decision problem is in \textsf{NP} when checking whether an instance belongs to the `Yes' subset can be performed in polynomial time.\footnote{The name \textsf{NP} stems from the theoretical construct of a \emph{non-deterministic} Turing machine, which can explore multiple computational branches simultaneously.}
	\item[\textsf{NP}-hard:] A problem is \textsf{NP}-hard if every problem in \textsf{NP} can be \emph{reduced} to it in polynomial time. 
	Intuitively, these problems are at least as hard as the hardest problems in \textsf{NP}.
	The \emph{halting problem}, about deciding whether a program will halt on an input, is \textsf{NP}-hard; remarkably, the halting problem, being undecidable,  lies outside of \textsf{NP}.
	\item[\textsf{NP}-complete:] This is the subset of problems that are both in \textsf{NP} and \textsf{NP}-hard. 
	They are the \emph{hardest} problems in \textsf{NP} in the sense that if any one of them were solvable in polynomial time, then every problem in \textsf{NP} would be. 
	A prominent example of \textsf{NP}-complete problem is the \emph{boolean satisfiability problem} (\textsf{SAT}), a statement proven in the Cook–Levin theorem \cite{arora_barak_2009}.
	Other classic examples include the \emph{traveling salesperson problem}, of determining whether there exists a path of length less than $l$ passing through $n$ points exactly once, or the \emph{subset sum problem} of finding a subset of numbers, among given $n$ integers, such that their sum is at most $s$.
	Both the problems can indeed be solved by exponential time algorithms: the \emph{traveling salesperson problem} can be solved in time complexity of (at most) $T(n) \sim n! \sim n^n$, and the \emph{subset sum problem} in time complexity $T(n) \sim 2^n$.
	However, they can both be verified, given a solution, in polynomial time, and they can both be reduced to the aforementioned \textsf{SAT}-problem, thus setting them in the class of \textsf{NP}-complete problems.
\end{description}
Figure~\ref{Fig:Compl_Classes} illustrates the relationships among some of these fundamental complexity classes. We emphasize that these classes are not exhaustive; the computational complexity literature defines many additional classes for finer distinctions \cite{arora_barak_2009}. Furthermore, these classifications are not immutable: problems currently believed to lie in \textsf{NP} may one day be shown to belong to \textsf{P}, should new, efficient algorithms be discovered.

In light of these definitions, we can restate the central computational questions motivating this work: 
\begin{claimbox} 
	Given an effective field theory with unknown UV completion, does the problem of \emph{constructing} the convex hull generated by some superextremal particles, and \emph{verifying} the Weak Gravity Conjecture lie in the class \textsf{P}? And within a theory that satisfies the conjecture, is the problem of \emph{finding} the minimal states that realize the conjecture solvable in polynomial time? 
\end{claimbox}
We will tackle the first question in Section~\ref{sec:WGCr}, and turn to the second problem in Section~\ref{sec:WGCm}.

\section{Realizations of the Weak Gravity Conjecture}
\label{sec:WGCr}

In this section, we introduce the \textsf{WGCHull} algorithm, a tool designed to construct the convex hull generated by a set of superextremal particles, and test the Weak Gravity Conjecture within effective field theories of the form \eqref{eq:WGC_S2}. 
Using this algorithm, we will estimate the computational complexity of checking the Weak Gravity Conjecture and analyze how it scales with the number of $U(1)$ gauge fields, denoted $N$.

\subsection{The \textsf{WGCHull} algorithm: the setup}
\label{sec:WGCr_intro}

To construct the algorithm, we must first specify its input data and the assumptions underpinning its implementation.

Consider an effective field theory governed by an action of the form \eqref{eq:WGC_S2}.
The first piece of the input is the number of independent abelian gauge fields $N$ that dynamically enter the action \eqref{eq:WGC_S2}.
A second piece of the input data is the black hole extremality sphere.
If the low-energy dynamics of the effective theory are entirely captured by the terms appearing in \eqref{eq:WGC_S2}, then the black hole extremality sphere is most readily given by \eqref{eq:WGC_BHextr_b}. 
If additional terms participate to the low-energy dynamics -- most notably, massless, or light scalar fields -- then the black hole subextremality region may differ, with its boundary not being a sphere, and has to be inferred from explicit black hole solutions of the action \eqref{eq:WGC_S2}. 
In particular, if the couplings of the gauge fields with the scalar fields are of \emph{dilatonic} kind  -- as for instance, in \cite{Gibbons:1987ps,Garfinkle:1990qj,Horowitz:1991cd,Heidenreich:2015nta} -- then the extremality bound \eqref{eq:WGC_BHextr_b} may change because the numerical factor on the right hand side differs.

More generically, whenever the scalar field space is more complicated, the black hole extremality vectors ${\bf Z}_{\rm ext}$ may draw an ellipsoid, rather than a sphere -- see, for example, \cite{Gendler:2020dfp,Bastian:2020egp}.
In the following, as a simplification assumption, we will assume that extremal black holes always describe a sphere in the charge-to-mass vectors ${\bf z}$-space; more involved cases may be reduced to this one after appropriate rescalings of the charge-to-mass ratio vectors ${\bf z}$.

\begin{wrapfigure}{r}{0.33\textwidth}
	\centering
	\includegraphics[width=0.30\textwidth]{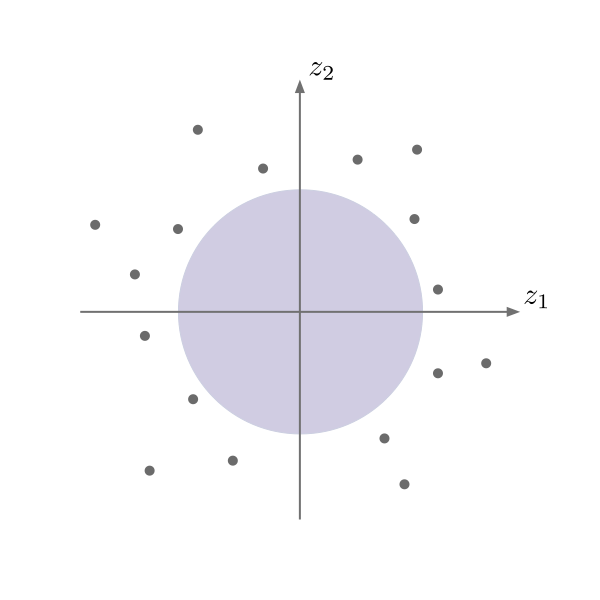}
	\caption{\footnotesize A two-dimensional depiction of the input data.
		\label{Fig:WGCr_Input}}
\end{wrapfigure}

Furthermore, we assume knowledge of a set of $N_s$ species of superextremal species of particles. 
As in Section~\ref{sec:WGC_CH}, we label their charge-to-mass ratio vectors ${\bf z}^{(i)}$, with $i = 1, \ldots, N_s$, obeying $|{\bf z}^{(i)}| > \alpha$, and they are part of the set $\mathcal{S} = \{ {\bf z}^{(i)}\; :\; i = 1, \ldots, N_s \}$.\footnote{As discussed in Section~\ref{sec:WGC_CH}, if the Weak Gravity Conjecture is realized, the final hull ought to trivially contain the origin. Here, we assume this is achieved through linear combinations of the ${\bf z}^{(i)}$, without explicitly including the origin in the set. Indeed, if the origin is not included, then $\mathcal{S}$ cannot realize the Weak Gravity Conjecture in the first place.}
To form an $N$-dimensional convex polytope with some of the vectors ${\bf z}^{(i)}$ determining its vertices, we must have $N_s \geq N + 1$. 
In fact, the charge-to-mass ratio vectors ${\bf z}^{(i)}$ have to be in enough number to ensure that a closed $N$-dimensional polyhedron can be constructed in the first place. 
In $N$ dimensions, the simplest polyhedron is an \emph{$N$-simplex}, with $N+1$ vertices (for instance, in two dimensions, $2$-simplices are triangle, or in three-dimensions $3$-simplices are tetrahedra).

Thus, the full input to the algorithm consists of: the number of $U(1)$ gauge fields $N$, the black hole extremality sphere in the ${\bf z}$-space, and the set $\mathcal{S}$ of $N_s$ superextremal charge-to-mass vectors ${\bf z}^{(i)}$.
In the two-dimensional case, the input data is of the form depicted in Figure~\ref{Fig:WGCr_Input}, where we plot the endpoints of the charge-to-mass ratio vectors ${\bf z}^{(i)}$ in $\mathcal{S}$. 
Given these ingredients, the \textsf{WGCHull} algorithm determines whether the theory satisfies the Weak Gravity Conjecture, and which are the particles that determine the hull realizing it. 
Formally, its structure is summarized as follows:
\begin{center}
	\begin{tabular}{ l l }
		\textbf{\textsf{WGCHull}} & \\ 
		\hline
		\textcolor{colorloc1}{\textsf{Question:}} & \textit{Does the effective theory obey the Weak Gravity Conjecture?}  \\ 
		\textcolor{colorloc2}{\textsf{Input:}}  & number of gauge fields $N$;
		\\
		&black hole extremality sphere; 
		\\
		& set of superextremal particles $\mathcal{S}$, with charge-to-mass ratio vectors ${\bf z}^{(i)}$ \\  
		\textcolor{colorloc3}{\textsf{Output:}}  &  Convex Hull $C$; Yes/No
	\end{tabular}
\end{center}
The algorithm proceeds in two steps: first, a subalgorithm outputs the (maximal) convex hull that the set of superextremal particles $S$ construct; then, a second subalgorithm verifies whether the convex hull so constructed contains the black hole extremality sphere.

The time complexity of the \textsf{WGCHull} algorithm will be expressed as the number of steps that the algorithm needs to halt as a function of the number of gauge fields $N$, and the number of superextremal particles $N_s$ in the set $\mathcal{S}$. 
Indeed, the time complexity of the \textsf{WGCHull} algorithm is the sum of the time complexities of the two subroutines that we describe below.

\subsection{Step I: construction of the convex hull}
\label{sec:WGCr_CH}

The first component of our algorithm is a subroutine that constructs the convex hull containing the set of superextremal particles $\mathcal{S}$. 
We call this subroutine \textsf{WGCHull (I)}.

\noindent\textbf{The \textsf{WGCHull (I)} algorithm.} Schematically, this first subroutine is structured as follows:
\begin{center}
	\begin{tabular}{ l l }
		\textbf{\textsf{WGCHull (I)}} & \\ 
		\hline
		\textcolor{colorloc1}{\textsf{Question:}} & \textit{What is the convex hull $C$ that contains the set of superextremal particles $\mathcal{S}$?}  \\ 
		\textcolor{colorloc2}{\textsf{Input:}}  & number of gauge fields $N$;
		\\
		& set of superextremal particles $\mathcal{S}$, with charge-to-mass ratio vectors ${\bf z}^{(i)}$ \\  
		\textcolor{colorloc3}{\textsf{Output:}}  &  Convex hull $C$ containing the set of superextremal particles $\mathcal{S}$
	\end{tabular}
\end{center}
Several algorithms are available to construct convex hulls containing a given set of points -- see, for instance,~\cite{CompGeo} for a review on the subject and references.
We will employ the \textsf{Quickhull} algorithm introduced in \cite{Greenfield1990APF,BDH}. 
While not being the most efficient algorithm (see, in particular, \cite{Chazelle1993}), it can be easily implemented to construct convex hulls in arbitrary dimensions.

\begin{figure}[t]
	\centering
	\includegraphics[width=0.6\textwidth]{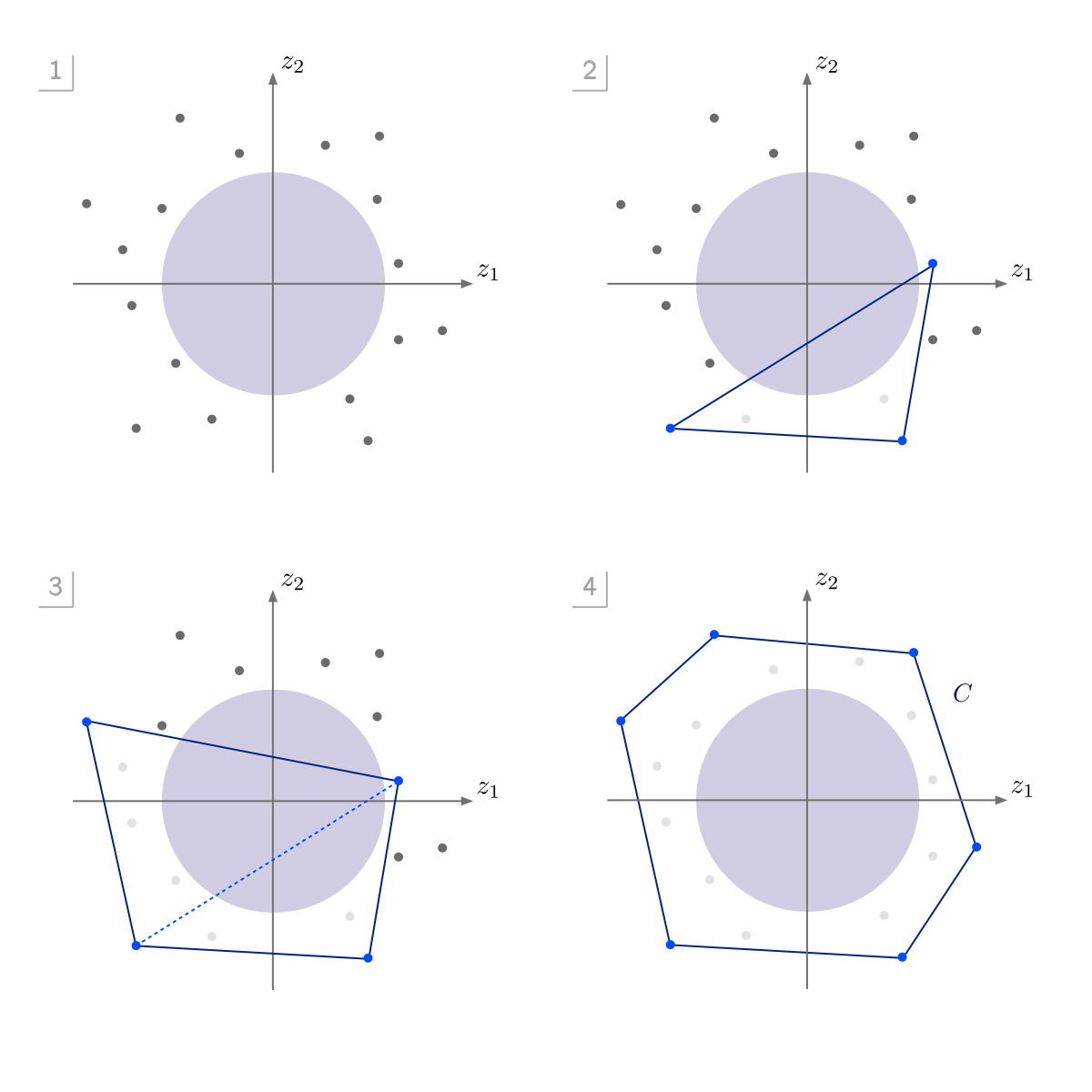}
	\caption{\footnotesize The idea behind the iterative construction of the convex hull proposed by the \textsf{Quickhull} algorithm.
		\label{Fig:WGCr_Hull_id}}
\end{figure}

Here, for the ease of exposition, we review the main idea behind the \textsf{Quickhull} algorithm; we refer to Appendix~\ref{sec:Hull_gen} for a detailed explanation of the algorithm in the framework of the effective field theories introduced in Section~\ref{sec:WGC_CH}.
Consider the set $\mathcal{S}$ of superextremal particles, with charge-to-mass ratio vectors ${\bf z}^{(i)}$ lying outside the black hole extremality sphere.
The convex hull, expressed in the $\mathcal{H}$-representation \eqref{eq:WGC_conv_Hrep}, that the vectors ${\bf z}^{(i)}$ generate can be obtained via an iterative procedure, consisting of determining, one by one, the \emph{facets} of the final hull.
Indeed, such facets can be described via hyperplanes of the form \eqref{eq:WGC_conv_cHa}.
One could then start with a random hull that only a handful of points in the set $\mathcal{S}$ generate.
Then, iteratively, one can consider points in the set $\mathcal{S}$ that are \emph{external} to the hull so constructed, and adjust the facets of the hull so that these points are included, either as vertices of the new hull, or as internal points.
In Figure~\ref{Fig:WGCr_Hull_id} is a schematic depiction of the algorithm in the case where the number of independent gauge fields is just two.

In sum, the subroutine \textsf{WGCHull (I)} delivers the convex hull $C$ containing all the superextremal particles in the set $\mathcal{S}$ via a recursive procedure. 
Furthermore, the output convex hull $C$ is expressed in $\mathcal{H}$-representation: it is defined by a collection of half-planes, and associated linear inequalities, corresponding to its facets.
However, we stress that, due to how the algorithm is structured, it is easy to also keep track of the vertices of the hull throughout, and store them.
Thus, the \textsf{WGCHull (I)} further provides information about the superextremal particles -- specifically, those whose charge-to-mass ratio vectors determine the vertices of the hull -- that are responsible for the creation of the final hull.

\noindent\textbf{Computational complexity of the \textsf{WGCHull (I)}.}  The computational complexity of the general \textsf{Quickhull} algorithm has been obtained in \cite{BDH}, a result which we now review.
As stressed above, the \textsf{Quickhull} algorithm is critically based on the iterative construction of the facets of the final hull.
However, the number of facets presents a computational bottleneck: given $n$ vertices, it can be shown that the number of facets $f_n$ of the hull constructed in $\mathbb{R}^N$ out of those vertices is bounded as\footnote{Given a positive, real number $a$, the floor function $\lfloor a \rfloor$ gives the greatest integer less than or equal to $a$.}
\begin{equation}
	\label{eq:QH_Compl_fr_bound}
	f_n \leq 2 \sum\limits_{k = 0}^{\lfloor \frac N2 \rfloor} \begin{pmatrix} n \\ k \end{pmatrix}
\end{equation}
as explained, for instance, in \cite{Ziegler2012}.
In particular, for large $n$, with $n \geq N + 1$, the maximum number of facets $f_n^{\rm max}$ behaves as
\begin{equation}
	\label{eq:QH_Compl_fr_est}
	f_n^{\rm max} = \mathcal{O} \left( n^{\lfloor \frac N2 \rfloor} / \left\lfloor \frac N2 \right\rfloor! \right) \,.
\end{equation}
Clearly, in the problem under consideration, the worst case occurs when the superextremal particles in $\mathcal{S}$ determine the vertices of the final hull, meaning $n = N_s$. 
Employing this observation, in~\cite{BDH} it is shown that the time complexity of the general \textsf{Quickhull} algorithm, and whence of the \textsf{WGCHull (I)} algorithm is
\begin{equation}
	\label{eq:QH_Compl_WGCHull_I}
		T_{\textsf{WGCHull (I)}}(N,N_s) \sim \mathcal{O} \left( \frac{N_s f_r}{r} \right) \sim \mathcal{O} \left( \frac{N_s r^{\lfloor \frac N2 \rfloor - 1}}{\lfloor \frac N2 \rfloor! } \right) \,,
\end{equation}
which holds for $N \geq 0$, and where $r \leq N_s$ is the number of \emph{processed points}; in the worst case, $r = N_s$, and $T(N,N_s) \sim \mathcal{O} \left(N_s^{ \lfloor \frac N2 \rfloor} \right)$.

Thus, we see that, if we assume $N$ to be fixed, the computational complexity of the \textsf{WGCHull (I)} algorithm is polynomial in the number of superextremal particles $N_s$.
However, since the maximum number of facets \eqref{eq:QH_Compl_fr_est} scales \emph{exponentially} with the number of gauge fields $N$, in general, the computational complexity of the \textsf{WGCHull (I)} algorithm is exponential in the input size.

\subsection{Step II: determining whether the convex hull covers the extremality sphere}
\label{sec:WGCr_cov}

The convex hull $C$ returned by \textsf{WGCHull (I)} contains all superextremal particles in $\mathcal{S}$, but this alone does not ensure that it satisfies the Convex Hull Weak Gravity Conjecture. 
Indeed, we must determine whether $C$ also contains the black hole extremality sphere.
This verification is performed by a second subroutine, \textsf{WGCHull (II)}, which we now illustrate.

\begin{figure}[t]
	\centering
	\includegraphics[width=0.9\textwidth]{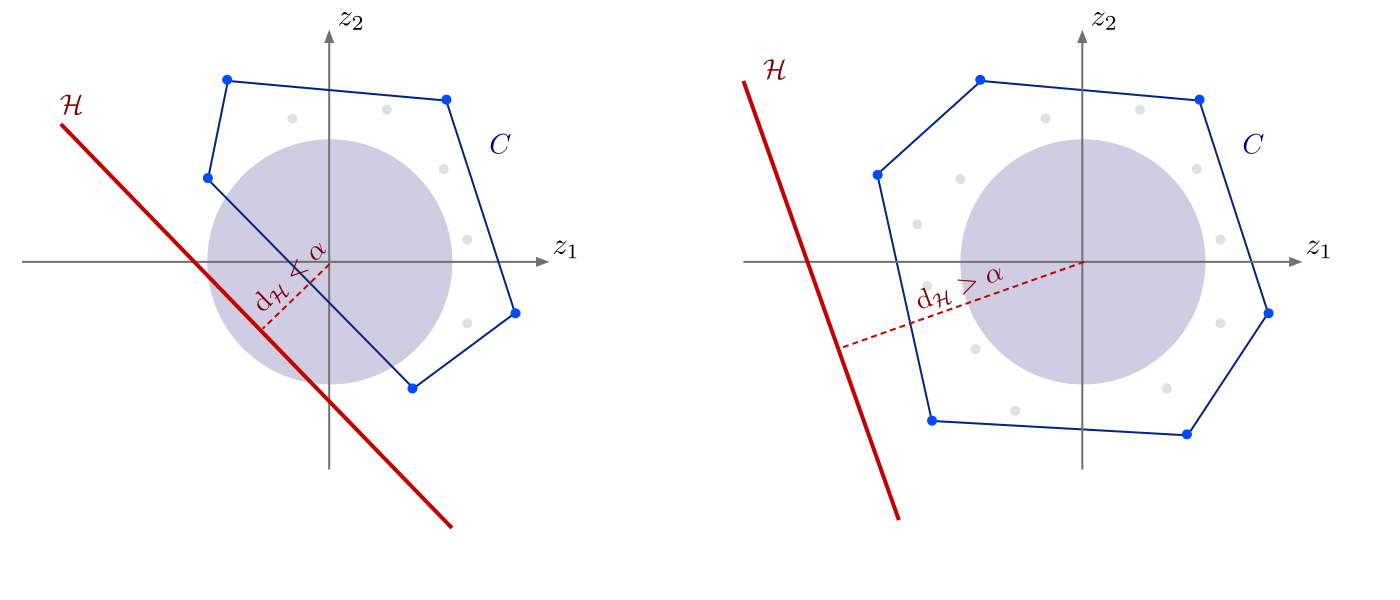}
	\caption{\footnotesize On the left, a two-dimensional example of convex hull $C$ not covering the black hole extremality disk, with hyperplane $\mathcal{H}$ with all the vertices of $C$ above it crossing the extremality disk; conversely, on the right, a convex hull covering the extremality disk, with hyperplane $\mathcal{H}$ with all the vertices of $C$ above it, not crossing the extremality disk.
		\label{Fig:WGCr_Plane}}
\end{figure}

\noindent\textbf{The \textsf{WGCHull (II)} algorithm.} This second subalgorithm is structured as follows:
\begin{center}
	\begin{tabular}{ l l }
		\textbf{\textsf{WGCHull (II)}} & \\ 
		\hline
		\textcolor{colorloc1}{\textsf{Question:}} & \textit{Does the convex hull $C$ contain the black hole extremality sphere?}  \\ 
		\textcolor{colorloc2}{\textsf{Input:}}  & number of gauge fields $N$;
		\\ & black hole extremality sphere; 
		\\
		& the convex Hull $C$ constructed by \textsf{WGCHull (I)}\\  
		\textcolor{colorloc3}{\textsf{Output:}}  &  Yes/No
	\end{tabular}
\end{center}
The idea, depicted in Figure~\ref{Fig:WGCr_Plane}, is the following: if the convex hull $C$ fails to enclose the extremality sphere, there exists a hyperplane $\mathcal{H}$ such that all vertices of $C$ lie above it and $\mathcal{H}$ intersects the sphere, lying at a distance ${\rm d}_{\mathcal{H}} < \alpha$ from the origin; conversely, if $C$ does contain the sphere, then any such hyperplane must lie at a distance ${\rm d}_{\mathcal{H}} > \alpha$ from the origin.
Therefore, in order to check whether the convex hull $C$ that the subalgorithm \textsf{WGCHull (I)} outputs, it is enough to find the hyperplane $\mathcal{H}$ with all the vertices of $C$ lying above it, and closest to the origin.  

\noindent\textbf{The steps of the \textsf{WGCHull (II)} algorithm.} Thus, the \textsf{WGCHull (II)} subalgorithm proceeds along the following steps:
\begin{enumerate}[itemsep=0mm, wide, labelwidth=!, labelindent=0pt]
	\specialitem{colorloc4}{1} Extract set of vertices $\mathcal{V}$ of the convex hull $C$, stored by the \textsf{WGCHull (I)} subroutine.
	\specialitem{colorloc4}{2} Define an hyperplane 
	\begin{equation}
		\label{eq:QH_II_hyp}
		\mathcal{H}_{\rm v}: \;{\bf a}^T_{\rm v} {\bf z} + 1 = 0\,,
	\end{equation}
	such that all the points in $\mathcal{V}$ lie above it:
	\begin{equation}
		\label{eq:QH_II_hyp_constr}
		\sigma_{\mathcal{H}_{\rm v}} ( {\bf z}^{(i)}_{\rm v} ) = + 1 \quad \text{for all} \;  {\bf z}^{(i)}_{\rm v} \in \mathcal{V},
	\end{equation}
	where, analogously to \eqref{eq:QH_II_hyp}, we have introduced the function $\sigma_{\mathcal{H}_{\rm v}} ({\bf z}) = {\rm sgn} ({\bf a}^T_{\rm v} {\bf z} + 1)$. 
	\specialitem{colorloc4}{3} Search for the hyperplane $\mathcal{H}_{\rm v}$ that is closest to the origin, minimizing its distance from the origin
	\begin{equation}
		\label{eq:QH_II_dist}
		{\rm d}_{\mathcal{H}_{\rm v}} = \frac{1}{\| {\bf a}_{\rm v} \|} \,.
	\end{equation}
	\specialitem{colorloc4}{4}  Check whether ${\rm d}_{\mathcal{H}_{\rm v}} < \alpha$ or ${\rm d}_{\mathcal{H}_{\rm v}} \geq \alpha$.
	In the former case, the convex hull $C$ does not contain the black hole extremality sphere, and the algorithm outputs `No'; in the latter case, the convex hull $C$ does contain the black hole extremality sphere, and the algorithm outputs `Yes'. 
\end{enumerate}

\noindent\textbf{Computational complexity of \textsf{WGCHull (II)}.} 
In order to find the computational complexity of the \textsf{WGCHull (II)} subroutine, it is worth noticing that the algorithm can be mapped to a \textit{linearly constrained quadratic programming} problem \cite{Boyd_Vandenberghe_2004}:
\begin{center}
	\begin{tabular}{ l l }
		\multicolumn{2}{l}{\textbf{\textsf{Linearly constrained quadratic programming}}}  \\  
		\hline
		\textcolor{colorloc1}{\textsf{Question:}} & \textit{Maximize the quantity $Q = \frac12 {\bf x}^T P {\bf x} + {\bf q}^T {\bf x}$} \rule{0pt}{5mm}  \\
		& \textit{for ${\bf x} \in \mathbb{R}^n$, with ${\bf q} \in \mathbb{R}^n$, $P$ $n \times n$ positive definite matrix} \\ 
		& \textit{subject to the constraints ${\bf a}^T_\alpha {\bf x} \leq c_\alpha$, for some ${\bf a}_\alpha \in \mathbb{R}^n$, $c_\alpha \in \mathbb{R}$, $\alpha = 1, \ldots, m$}\\
		\textcolor{colorloc2}{\textsf{Input:}}  & $n \times n$ positive definite matrix $A$, ${\bf q}, {\bf a}_\alpha \in \mathbb{R}^n$, $c_\alpha \in \mathbb{R}$
		\\
		\textcolor{colorloc3}{\textsf{Output:}}  & vector ${\bf x}_{\rm max}$ for which $Q$ acquires its maximum values.
	\end{tabular}
\end{center}

While being considered \textsf{NP} for long time, in case of a positive-definite matrix $P$, the problem has been shown to be solvable in polynomial time.
Indeed, developing on earlier results on linear programming based on the usage of the \emph{ellipsoid method} \cite{papadimitriou82}, in \cite{Ye1989} it was shown that the time complexity of the above quadratic programming problem is
\begin{equation}
	\label{eq:QH_Compl_WGCHull_II0}
	T(n) = \mathcal{O} (L n^3 )
\end{equation}
where $n$ is the dimension of the vector space where the vector ${\bf x}$ resides, and $L$ is the number of \emph{bits} in the input, namely
\begin{equation}
	\label{eq:QH_Compl_WGCHull_L}
	L \sim n^2 + nm + [ \log |P|] \,,
\end{equation}
where $P$ is the product of all nonzero coefficients appearing in the matrix $P$, in the vectors ${\bf q}$, ${\bf a_\alpha}$, and the nonzero $c_\alpha$. 

Let us now specialize to the \textsf{WGCHull (II)} case: $n$ gets identified with the number of independent gauge fields $N$, and $m$ is the number of vertices of the hull generated by \textsf{WGCHull (II)}, and is a fraction of $N_s$; furthermore, $P$ is proportional to the identity matrix, while ${\bf q} = 0$, $c_\alpha = 1$; finally ${\bf a}_\alpha$ contain, at most, $N N_s$ components.
Thus, in our case, $L \sim N^2 + N N_s + [\log N N_s]$.
Hence,  the time complexity of the subalgorithm \textsf{WGCHull (II)} can be estimated as
\begin{equation}
	\label{eq:QH_Compl_WGCHull_II}
	T_{\textsf{WGCHull (II)}}(N,N_s ) \sim \mathcal{O} (L N^3) \sim \mathcal{O} (N_s N^4) \,,
\end{equation}
displaying that it can deliver an output in polynomial time in the number of superextremal particles, and number of gauge fields $N$.

\subsection{The computational complexity of the \textsf{WGCHull} algorithm}
\label{sec:WGCr_compl}

We can finally analyze the computational resources required by a Turing machine to obtain the constructions predicted by the Convex Hull Weak Gravity Conjecture. Indeed, the time computational complexity of \textsf{WGCHull} is the sum of the two computational complexities \eqref{eq:QH_Compl_WGCHull_I}, \eqref{eq:QH_Compl_WGCHull_II} of its two constituting subalgorithms:
\begin{equation}
	\label{eq:QH_Compl_WGCHull}
	T_{\textsf{WGCHull}}(N,N_s ) =  T_{\textsf{WGCHull (I)}}(N,N_s ) + T_{\textsf{WGCHull (II)}}(N,N_s )\,.
\end{equation}
In particular, we observe, as the number of gauge field $N$ increases, the time complexity $T_{\textsf{WGCHull}}(N,N_s )$ increases exponentially with $N$, due to the \textsf{WGCHull (I)} algorithm not performing well, as indicated by \eqref{eq:QH_Compl_WGCHull_I}.
We thus arrive at the following conclusion:

\begin{claimbox}
	Consider a given effective field theory, for which the black hole subextremality region is a sphere in the charge-to-mass ratio vector space, and supporting a set of superextremal particles. 
	The problem of constructing the convex hull in the ${\bf z}$-space generated by the superextremal particles, and then checking the Convex Hull Weak Gravity Conjecture resides, in general, in the class \textsf{EXP}, meaning that a Turing machine requires a time that is exponential in the number of gauge fields to deliver an output.
\end{claimbox}

As stressed in Section~\ref{sec:WGC_compl}, the computational complexity is a characteristic of the problem, not of the specific algorithm employed.
Hence, the estimate \eqref{eq:QH_Compl_WGCHull} ought to be considered as an upper bound on the effective time complexity of the problem.
Indeed, this does not preclude the possibility that, in some selected subcases, the computational complexity is lower than \eqref{eq:QH_Compl_WGCHull}.

\subsection{Improvements, limitations, and landscape scans}
\label{sec:WGCr_gen}

While the conclusion drawn in the previous section is general, here we comment on possible modifications of the \textsf{WGCHull} algorithm, in both structure and scopes, and investigate in which cases it offers efficient checks of the Convex Hull Weak Gravity Conjecture.

\noindent\textbf{The \textsf{WGCHull} algorithm for a small number of gauge fields.} If the number of gauge fields is fixed, and sufficiently small, the \textsf{WGCHull} algorithm may still be efficient in delivering the convex hull.
For instance, the two-dimensional version of the \textsf{WGCHull (I)} algorithm presented in Appendix~\ref{sec:Hull_gen2} displays the time complexity \eqref{eq:QH_Compl_I} that is polynomial in the sole input $N_s$; consequently, the total time complexity is also polynomial in $N_s$.
Thus, in these cases, realizations of the Convex Hull Weak Gravity Conjecture can be constructed in polynomial time.

\noindent\textbf{Bypassing the \textsf{WGCHull (I)} subroutine.} The \textsf{WGCHull (II)} subalgorithm can be applied directly to the set of superextremal particles $\mathcal{S}$, without the need of finding their convex hull in the first place, and thus bypassing the first \textsf{WGCHull (I)} subroutine.
In this way, the \textsf{WGCHull (II)} indeed tells whether a set of superextremal particles $\mathcal{S}$ can construct a hull such that it encloses the black hole extremality sphere.

Given the computational complexity estimate \eqref{eq:QH_Compl_WGCHull_II} of \textsf{WGCHull (II)}, this simpler question can be addressed by a Turing machine in a time that is polynomial in both the number of gauge fields $N$, and the number of superextremal particles $N_s$.

It is however worth remarking that the answer delivered by the \textsf{WGCHull (II)} subroutine alone may not be informative enough: in fact, \textsf{WGCHull (II)} does not give any information about how to construct the hull, or which of the superextremal particles is necessary, and which is redundant to construct the convex hull. 
As such, the sole application of the \textsf{WGCHull (II)} subroutine, while useful for an exploratory check of the realization of the Convex Hull Weak Gravity Conjecture, may not be conceptually, or experimentally appealing.

\noindent\textbf{Including moduli dependences.} In Section~\ref{sec:WGCr_intro} it was assumed that the effective field theory that the \textsf{WGCHull} algorithm tests is devoid of a moduli space.
This assumption is quite restrictive in perspective of an ultraviolet completion of the effective theory within string theory.
Indeed, generically, several scalar fields appear in the effective field theory, and they actively participate in the realization of the Weak Gravity Conjecture \cite{Palti:2017elp}.
In fact, within the action \eqref{eq:WGC_S2}, the gauge-kinetic matrix $f_{IJ}$ may get promoted to a moduli-dependent matrix.
Consequently, the black hole subextremality region, whose boundary is determined requiring the black hole horizons to collapse into a single one, may depend on the values of the scalar fields, and might not be a simple sphere as in \eqref{eq:WGC_BHextr_b}.
For example, in \cite{Gendler:2020dfp,Bastian:2020egp}, it has been shown that for some typical scalar field couplings, the black hole extremality vectors ${\bf Z}_{\rm ext}$ draw an ellipsoid, rather than a sphere.
However, if such is the case, it is enough to rescale the charge-to-mass ratio vectors ${\bf z}$ so that the black hole subextremality region is described by a sphere as in \eqref{eq:WGC_BHextr_b}.
Then, the \textsf{WGCHull} algorithm can be applied with no further modification, and ought to be capable of performing efficient checks of Weak Gravity Conjecture for small number of gauge fields $N$.

\noindent\textbf{Unknown extremality region.} 
As mentioned above, the identification of black hole extremal solutions passes through the solution of the equations of motion of the action \eqref{eq:WGC_S2}, employing an appropriate ansatz for the spacetime metric. 
These solitonic solutions dictate how \emph{all} the fields entering the action \eqref{eq:WGC_S2} vary across the spacetime.
In particular, if scalar fields are present in the low-energy effective field theory, they may appear with highly non-trivial, non-canonically normalized kinetic terms, and they may enter the gauge kinetic matrix $f_{IJ}$ in complicated ways. 
Indeed, such couplings may be so cumbersome that they could make very hard to find explicit extremal solitonic solutions.

In order to overcome such a difficulty, one could try to find only some particular extremal, or subextremal solutions -- for instance, by fixing the scalar fields at some given values.
Let us call $\mathcal{E}$ the set of isolated charge-to-mass ratio vectors ${\bf z}^{(l)}$, $l = 1, \ldots, L$, of the particular solutions found in this way.
Starting from the set $\mathcal{E}$, one could grasp what is the extremality region of the black hole.

For instance, by employing the \textsf{WGCHull (I)} algorithm to the set of extremal and subextremal black holes $\mathcal{E}$, one could find the convex hull spanned by the vectors ${\bf z}^{(l)}$.
Indeed, the black hole subextremality region ought to contain such a convex hull.
One could then check whether the convex hull of the extremal and subextremal black holes $\mathcal{E}$ is contained in the convex hull generated by the superextremal particles in the set $\mathcal{S}$.
In principle, this can be achieved by using polytope-containment algorithms, such as those of
\cite{Eaves1982,Freund1985}. 
However, since the black hole subextremality region is just approximated by a convex hull, such a procedure can only be used as first, informative step as a validation for the Weak Gravity Conjecture.

Alternatively, starting from the set of extremal and subextremal black holes $\mathcal{E}$, one could first find the smallest sphere enclosing them as an approximation for the black hole subextremality region. 
This can be achieved by exploiting, for instance, the polynomial-time algorithms proposed in  \cite{Welzl1991,Chan2018}.
Then the \textsf{WGCHull} algorithm can be applied by considering the sphere so obtained as input for its second subroutine.

\noindent\textbf{Scan over many effective field theories.} The \textsf{WGCHull} algorithm offers a tool to test the Convex Hull formulation of the Weak Gravity Conjecture in a single, given effective field theory.

Now, suppose we are given a large number of effective theories, for which we know the precise shape of the black hole subextremality region, and the spectrum of superextremal particles that could potentially realize the Weak Gravity Conjecture. 
In this scenario, we could deploy \textsf{WGCHull} to each theory individually, verifying in turn whether the corresponding convex hull of charge-to-mass ratios satisfies the conjecture.

However, a significant obstacle arises: the number of $U(1)$ gauge fields in typical landscape theories can vary widely -- and in some cases, reach into the hundreds. Constructing the full convex hull for each theory in such high-dimensional charge spaces is computationally prohibitive.

A more tractable approach is to relax the problem: rather than explicitly building the convex hull, we simply ask whether the known set of superextremal particles could in principle generate a hull that covers the black hole extremality sphere. 
This question can be answered efficiently using the \textsf{WGCHull (II)} subroutine alone.
Provided the number of theories under consideration is not itself exponentially large, this reduced criterion allows us to systematically scan a subset of the landscape and flag theories where no realization of the Weak Gravity Conjecture is possible.

\section{Minimal realizations of the Weak Gravity Conjecture}
\label{sec:WGCm}

The convex hull obtained via the \textsf{WGCHull (I)} subroutine (see Section~\ref{sec:WGCr_CH}) is not necessarily minimal. 
This means that, even if the hull satisfies the Convex Hull Weak Gravity Conjecture, it may include superextremal particles that are not strictly necessary for this purpose. 
Indeed, the algorithm ensures that the convex hull encompasses all input particles, and the resulting vertices then define the charge-to-mass ratio vectors that fulfill the conjecture in the sense of~\cite{Cheung:2014vva}.

Yet, this approach can yield a redundant convex hull, larger than required. 
This typically arises when the input set $\mathcal{S}$ is a finite sample from a broader structure, such as a sublattice~\cite{Heidenreich:2016aqi} or an infinite tower, along some or all the allowed charge directions~\cite{Heidenreich:2015nta,Andriolo:2018lvp,FierroCota:2023bsp}. 
In these cases, the vertices of the convex hull tend to align with particles possessing the largest charge-to-mass ratios in the set.
This motivates the search for minimal realizations of the Convex Hull Weak Gravity Conjecture: the smallest subset of superextremal particles whose charge-to-mass ratio vectors suffice to cover the entire black hole subextremality region.

Understanding such minimal realizations is crucial for any concrete test of the conjecture. 
Indeed, suppose we have theoretical or experimental evidence for a set of superextremal particles  $\mathcal{S}_{\rm exp}$.
These may not be enough to determine a convex hull that fully covers the black hole extremality sphere; in such a case, running the \textsf{WGCHull} algorithm would deliver a negative answer.
Then, one strategy is to expand the set $\mathcal{S}_{\rm exp}$ to a new set $\mathcal{S}$ so as to include putative, new superextremal particles such that the new set $\mathcal{S}$ could create a convex hull covering the black hole subextremality region, and thus realizing the Weak Gravity Conjecture. 

The extension of the set $\mathcal{S}_{\rm exp}$ to the larger set $\mathcal{S}$ can be done in several ways.
While this extension could be done in \emph{trivial} ways, just by including new superextremal particles with very large charge-to-mass ratios, a more interesting question is to look for \emph{minimal} extensions of the set $\mathcal{S}_{\rm exp}$ in such a way that the conjecture is realized.

In the remainder of this section, we formalize the notion of `minimal realization' and investigate whether an algorithmic approach can efficiently identify such minimal sets.

\subsection{A first attempt: an enumeration algorithm}
\label{sec:WGCm_brute}

We begin with a simple, but conceptually clear method to look for minimal realization of the Weak Gravity Conjecture that relies on the \textsf{WGCHull} algorithm introduced in Section~\ref{sec:WGCr}, and based on an exhaustive search.
Here, we define a ‘minimal’ realization as follows: given a set of superextremal particles $\mathcal{S}$, the Convex Hull Weak Gravity Conjecture is minimally realized by a set of $n \leq N_s$ of particles in $\mathcal{S}$, if they realize the conjecture, and none of the sets with $n-1$ particles in $\mathcal{S}$ can realize it.
In other words, $n$ is the minimum number of particles within the set $\mathcal{S}$ whose convex hull can cover the black hole subextremality region.

In order to find these minimal realizations we can proceed by enumeration: firstly, we can list \emph{all} the subsets of superextremal particles that can be constructed from $\mathcal{S}$.
Clearly, some of these subsets cannot geometrically deliver a closed hull. 
As explained at the beginning of Appendix~\ref{sec:Hull_gend}, in $\mathbb{R}^N$ the simplest hull is a simplex, with $N+1$ vertices.
Hence, from the subsets of $\mathcal{S}$ we ought to exclude those with less than $N+1$ points, for they cannot construct a hull.

Then, one should apply the \textsf{WGCHull} on each of the subsets of $\mathcal{S}$ to single out which of them realize the Convex Hull Weak Gravity Conjecture.
The subsets that minimally realize the Convex Hull Conjecture are those with the smallest number of superextremal particles as vertices.
For instance, in $\mathbb{R}^N$, we may see whether all the allowed $N$-simplices, with $N+1$ vertices, can cover the black hole subextremality region.
If none of such simplices is capable of covering the subextremality region, we then pass to subsets of $N+2$ superextremal particles, and so on.

However, the enumeration procedure just proposed has a critical drawback.
Firstly, given the set $\mathcal{S}$ of $N_s$ superextremal particles, the number of subsets of $\mathcal{S}$ is $2^{N_s}$.
Among these, we  exclude the subset with $n \leq N$ points -- for large $N_s$, excluding these subsets does not change the scaling behavior of the total number of subsets in $N_s$.
Then, each of these subsets has to be fed to \textsf{WGCHull}; based on the estimations \eqref{eq:QH_Compl_WGCHull_I}, \eqref{eq:QH_Compl_WGCHull_II}, assuming $N$ to be fixed, for large $N_s$, the time complexity of the proposed enumeration procedure is
\begin{equation}
	T(N_s) = \mathcal{O} \left( N_s^c 2^{N_s} \right) \,,
\end{equation}
for some $c \in \mathbb{N}$.
Thus, the problem of listing all the subsets of superextremal particles $\mathcal{S}$, and single out those with minimum number of vertices realizing the Weak Gravity Conjecture is in the class \textsf{EXP}, even for small $N$.
This implies that a deterministic Turing machine would not halt in feasible time, and a more tractable strategy is needed.

\subsection{The search for a minimal convex hull}
\label{sec:WGCm_ch}

As noted earlier, the simplest object with which we can cover a ball in $\mathbb{R}^N$ is an $N$-simplex, characterized by $N+1$ vertices.
For instance, in $\mathbb{R}^2$, the simplest convex hull that covers a disk is a triangle, while in $\mathbb{R}^3$ the simplest polyhedron that can cover a ball is a tetrahedron.
Hence, we could look for such $N$-simplices, with $N+1$ of the $N_s$ superextremal particle species realizing it as its vertices.

But which $N$-simplex should we choose?
One natural criterion is to select the $N$-simplex of largest volume among those that are fully contained within the convex hull and encompass the subextremality region. 
Intuitively, a larger simplex has a higher chance of containing the subextremality region.

Momentarily, let us assume to have obtained the convex hull $C$ generated by a set of superextremal particles $\mathcal{S}$. 
Indeed, for low enough number of gauge fields $N$, the \textsf{WGCHull} algorithm can efficiently determine it.
As a further simplifying assumption, we suppose that the $N+1$ vertices of the $N$-simplex that we are looking for are also vertices of $C$.

This leads to the following algorithmic question:
\begin{center}
	\begin{tabular}{ l l }
		\multicolumn{2}{l}{\textbf{\textsf{Finding the largest $N$-simplex minimally realizing the Weak Gravity Conjecture}}}  \\  
		\hline
		\textcolor{colorloc1}{\textsf{Question:}} & \textit{What is the largest $N$-simplex containing the black hole subextremality region?} \rule{0pt}{5mm}  \\
		\textcolor{colorloc2}{\textsf{Input:}}  & Dimension of the charge-to-mass ratio vector space $N$
		\\
		& Convex hull $C$ generated by the set of superextremal particles $\mathcal{S}$\\
		\textcolor{colorloc3}{\textsf{Output:}}  & $N+1$ charge-to-mass ratio vectors (if they exist) generating the largest simplex 
		\\
		& containing the black hole subextremality region
	\end{tabular}
\end{center}

The problem of finding the largest $j$-simplex (with $j \leq N$) out of the vertices of a given convex hull has been studied in \cite{GKL}, and here we highlight the relevant results for our question.
Indeed, as a preliminary benchmark, in \cite{GKL}, it is suggested to simply list all the $N$-simplices that can be drawn out of the vertices of the convex hull.
Out of the $n$ vertices of the hull we can construct the following number of $N$-simplices
\begin{equation}
	N_{N\text{-simpl}} = \begin{pmatrix} n \\ N+1 \end{pmatrix} \sim \mathcal{O} \left( n^{N+1} \right) \,.
\end{equation}
In the worst case where the number of vertices coincide with the number of superextremal particles in the set $\mathcal{S}$, the number of $N$-simplices is $N_{N\text{-simpl}} \sim \mathcal{O} \left( N_s^{N+1} \right)$.
We can then draw conclusions similar to those of Section~\ref{sec:WGCr_compl}: the number of $N$-simplices increases exponentially with the dimension $N$; thus, performing such an enumeration, and subsequent volume computations would require an exponential amount of time.
This renders the proposed algorithm inefficient, with a Turing machine not halting after a reasonable number of steps.
Clearly, a similar estimate of computational complexity would have hold if we would have looked for the largest simplex realized with the set of superextremal particles $\mathcal{S}$, rather than looking for an $N$-simplex realized out of the vertices of the hull $C$.

Although here we have just mentioned the possibility of finding the largest $N$-simplex via an enumeration-based algorithm, in \cite{GKL} the problem is examined in deeper details.
Indeed, therein it is shown that for convex hulls expressed in the $\mathcal{V}$-representation, the problem of finding the largest $j$-simplex is \textsf{NP}-hard, with no algorithm solving it in an amount of steps that is polynomial with the dimension $N$.

Furthermore, for convex hulls expressed in the $\mathcal{H}$-representation -- as those most readily provided by the \textsf{WGCHull} algorithm -- it can be shown that the problem of finding the largest $N$-simplex of a hull is \textsf{NP}-complete.
This is achieved by mapping the problem to the well-known \textsf{3SAT} problem\footnote{This is one of the famous Karp's 21 \textsf{NP}-complete problems, and consists of checking whether a Boolean formula of the form $\bigwedge_i  (u_{i_1} \vee u_{i_2} \vee u_{i_3})$ is satisfied for some assignment of the variables $u_{i_k}$. The \textsf{NP}-completeness of the \textsf{3SAT} problem is part of the Cook-Levin theorem \cite{arora_barak_2009}.}, which is indeed \textsf{NP}-complete \cite{GKL,BGKV}. 
This renders the problem of finding the largest $N$-simplex with largest volume that would minimally realize the Convex Hull Weak Gravity Conjecture one the \emph{hardest} problems in the \textsf{NP}-class.

\section{Conclusions and outlook}
\label{sec:Conclusions}

In this work, we have investigated the computational complexity underlying the Convex Hull formulation of the Weak Gravity Conjecture. 
We showed that the time that a Turing machine takes to construct the relevant convex hull and determine whether it encloses the black hole extremality sphere grows exponentially with the number of gauge fields. 
As a result, constructing bottom-up realizations of the Convex Hull Weak Gravity Conjecture within effective field theories typically lies in the \textsf{EXP} complexity class.

We further demonstrated that identifying minimal realizations -- those involving the smallest possible number of superextremal particles -- is computationally hard and, in some formulations, \textsf{NP}-complete. 
These difficulties persist unless we forgo constructing the convex hull explicitly. In that case, a Turing machine can efficiently determine whether a given set of superextremal particles could, in principle, generate a hull that covers the black hole extremality sphere.

It would be valuable to implement the \textsf{WGCHull} algorithm introduced in Section~\ref{sec:WGCr} to empirically validate these computational bottlenecks and track how the runtime scales with the number of gauge fields. We leave such implementations and benchmarks to future investigations.

It is important to emphasize that our analysis extends beyond the Weak Gravity Conjecture and applies more broadly to other Swampland constraints that rely on analogous convex hull constructions. Notable examples include the hull-based approaches used to diagnose effective field theory breakdowns associated with the Distance Conjecture~\cite{Calderon-Infante:2020dhm,Etheredge:2023odp,Etheredge:2024tok,Grieco:2025bjy}, as well as the recently proposed AdS version of the Weak Gravity Conjecture~\cite{Lin:2025wfe}. We expect that the computational barriers to hull constructions identified in this work persist in these contexts as well, with no significant improvement.

Lastly, it is worth noting that the computational barriers identified in this study might be bypassed through machine learning techniques. In particular, decision problems arising in quantum gravity effective field theories -- such as the one considered here -- are expected to be learnable using neural network-based algorithms \cite{Lanza:2024mqp}. However, machine learning is inherently statistical and thus susceptible to error. 
For instance, given a candidate set of superextremal particles, a machine learning algorithm could assess whether the configuration is more or less likely to satisfy the Weak Gravity Conjecture, rather than offering a definitive answer. 
The intricate relationship between computability and learnability remains an open and fascinating direction, which we intend to pursue in future work.

\noindent{\textbf{Acknowledgments.}} I am deeply grateful to Fabian Ruehle for precious, valuable comments on the draft, and I would also like to thank Thomas Grimm, Lukas Kaufmann, Mick van Vliet, Timo Weigand, Alexander Westphal for useful discussions and comments on the manuscript.
This work is supported in part by Deutsche Forschungsgemeinschaft under Germany’s Excellence Strategy EXC 2121 Quantum Universe 390833306, by Deutsche Forschungsgemeinschaft through a German-Israeli Project Cooperation (DIP) grant ``Holography and the Swampland” and by Deutsche Forschungsgemeinschaft through the Collaborative Research Center 1624 ``Higher Structures, Moduli Spaces and Integrability”.

\appendix

\section{Constructing the convex hull}
\label{sec:Hull_gen}

In Section~\ref{sec:WGCr_CH} we presented the first subroutine of the \textsf{WGCHull} algorithm, consisting of constructing the convex hull generated by a set of superextremal particles. 
Therein, we stated that an efficient, and versatile algorithm to construct the convex hull is the \textsf{Quickhull} algorithm~\cite{Greenfield1990APF,BDH}.
In this section, we review the algorithm, while translating \cite{Greenfield1990APF,BDH} in the effective field theory-language of Section~\ref{sec:WGC_CH}. 

For the sake of clarity, we will first present a version of the \textsf{Quickhull} algorithm that applies to the case where we have only two independent gauge fields, namely $N = 2$~\cite{Greenfield1990APF}; 
then, we will illustrate the more general version of the \textsf{Quickhull} algorithm that applies to effective field theories endowed with a generic number of gauge fields~\cite{BDH}.

\subsection{Constructing the convex hull in two dimensions}
\label{sec:Hull_gen2}

\begin{figure}[t]
	\centering
	\includegraphics[width=\textwidth]{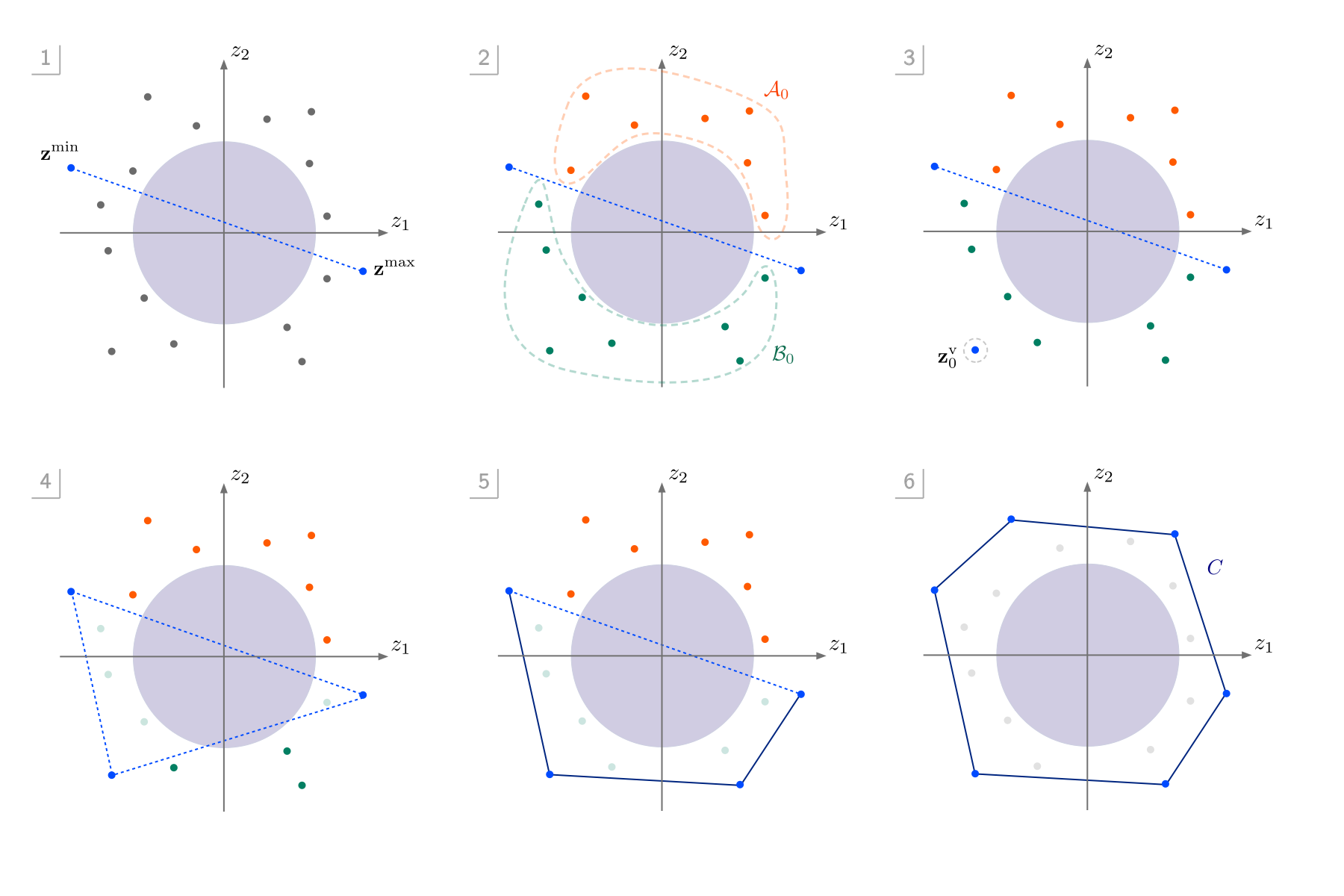}
	\caption{\footnotesize The steps of the two-dimensional version of the \textsf{Quickhull} algorithm.
		\label{Fig:WGCr_Hull}}
\end{figure}

Consider an effective theory whose low-energy dynamics are governed by just two $U(1)$ gauge fields, so that the space spanned by the charge-to-mass ratio vectors ${\bf z} = (z_1, z_2)$ is a two-dimensional plane. 

The two-dimensional \textsf{Quickhull} algorithm proceeds according to the following steps (see Figure~\ref{Fig:WGCr_Hull} for a visual representation):
\begin{enumerate}[itemsep=0mm, wide, labelwidth=!, labelindent=0pt]
	\specialitem{colorloc4}{1} Identify the two superextremal particles with minimum and maximum coordinate $z_1^{(i)}$ for their charge-to-mass ratio vectors ${\bf z}^{(i)}$; if several of them are present, the particles with minimum and maximum coordinate $z_1^{(i)}$ can be randomly picked. Denote with ${\bf z}^{\rm min}$ and ${\bf z}^{\rm max}$ the charge-to-mass ratio vectors of the particles so chosen, and draw a line between them. We describe such a line as ${\bf a}^T_0 {\bf z} + 1 = 0$, with ${\bf a}_0^T = (a_{0,1}, a_{0,2} )^T$. 
	\specialitem{colorloc4}{2} Introduce the function 
	\begin{equation}
		\label{eq:QH_sign_0}
		\sigma({\bf z}) = {\rm sgn} ({\bf a}^T_0 {\bf z} + 1) \,.
	\end{equation}
	For a superextremal particle with charge-to-mass vector ${\bf z}^{(i)}$ that lies above the line ${\bf a}^T_0 {\bf z} + 1$, $\sigma({\bf z}^{(i)}) = 1$; if it lies below,  $\sigma({\bf z}^{(i)}) = -1$.
	Employing \eqref{eq:QH_sign_0}, we split the set of the remaining superextremal particles into two sets $\mathcal{A}_0$, the subset of particles above the line ${\bf a}^T_0 {\bf z} + 1 = 0$, and $\mathcal{B}_0$, the subset of those below the line. 
	\specialitem{colorloc4}{3} Focus on one of the two subsets constructed above: for instance, in Figure~\ref{Fig:WGCr_Hull} we consider the subset $\mathcal{B}_0$ of the superextremal particles that lie below the line ${\bf a}^T_0 {\bf z} + 1 = 0$.
	For each of the particles within this subset, with charge-to-mass ratio vector ${\bf z}^{(i)}$, we compute its distance from the line via the standard formula
	\begin{equation}
		\label{eq:QH_d_0}
		{\rm d}({\bf z}^{(i)}) = \frac{\| {\bf a}^T_0 {\bf z}^{(i)} + 1  \|}{\| {\bf a}_0 \|}\,,
	\end{equation}
	with $\| \cdot \|$ denoting the standard Euclidean norm, namely $\| {\bf z} \|^2 = (z_1)^2 + (z_2)^2$.
	Then, we select one of the charge-to-mass ratio vectors, that we call ${\bf z}^{\rm v}_0$, located at maximum distance from the line.
	\specialitem{colorloc4}{4} Construct the triangle having the charge-to-mass ratios ${\bf z}^{\rm min}$, ${\bf z}^{\rm max}$, and ${\bf z}^{\rm v}_0$ as its vertices. 
	Let us introduce the lines ${\bf a}^T_{1,0} {\bf z} + 1 = 0$ (${\bf a}^T_{1,1} {\bf z} + 1 = 0$) passing through ${\bf z}^{\rm v}_0$ and  ${\bf z}^{\rm min}$ (${\bf z}^{\rm max}$).
	From now on, we will be neglecting the points inside such a triangle, for they cannot be vertices of the final hull.
	\specialitem{colorloc4}{5} For each of the newly constructed lines ${\bf a}^T_{1,0} {\bf z} + 1 = 0$, ${\bf a}^T_{1,1} {\bf z} + 1 = 0$, repeat the steps from 2 to 4.
	\specialitem{colorloc4}{6} Continue until the separation of points dictated by the lines constructed as in step 2 would only lead to empty sets. 
	The lines so obtained are the edges of the final convex hull $C$.
\end{enumerate}

The computational complexity of such a simplified version of the \textsf{Quickhull} algorithm can be easily estimated. 
In fact, consider one of the recursive steps of the \textsf{WGCHull (I)}. 
Some of the superextremal particles might have been already eliminated in the previous steps, for they are not necessary for the determination of the final hull, being internal.
Hence, we are generically left with $n \leq N_s$  number of points.
At each of the recursive steps of \textsf{WGCHull (I)}, the subalgorithm performs two operations: it separates the set of $n$ points into two subsets, according to whether they reside above or below a given line; then, in a given subset, it finds the furthest point from the line.

Consider the former operation: after the subdivision, a number of $a n$ points, with $a = \frac{p}{n}$, where $p \in \mathbb{N}$, $p \leq n$ fall into one of the two subsets, and $b n$ points, with $b = 1- a$, fall into the second.
Then, in a successive step, the algorithm will be iterated on the two subsets separately.
We can then write down a recursive relation to estimate the time complexity.
Let us denote with $T(n)$ the time complexity of a \emph{single step} of the subalgorithm \textsf{WGCHull (I)}; then, since in the successive step we need to apply the algorithm to the two sets of $a n$ and $b n$ points, the time complexity to recursively complete the former step is
\begin{equation}
	\label{eq:QH_Compl_I_0}
	T(n) = T(a n) + T (b n) + \ldots
\end{equation}

However, at each step of the \textsf{WGCHull (I)} subalgorithm, we also need to determine which is the point, in each of the subsets, that is the farthest from the line.
Since we have $n$ points, we have to compute $n$ signed distances from the given line.
This can be achieved with a number of computation proportional to the number of points $n$.
Thus, the time complexity of the each step is recursively expressed as
\begin{equation}
	\label{eq:QH_Compl_I_rec}
	T(n) = T(a n) + T (b n) + \mathcal{O}(n) \,, 
\end{equation}

Two extreme cases may occur.
The best case happens when, at each step of the iteration, the set of $n$ points is equally split, with $a = b = \frac12$. 
Then, the time complexity \eqref{eq:QH_Compl_I_rec} becomes
\begin{equation}
	\label{eq:QH_Compl_I_a}
	T(n) = 2 T \left(\frac n2\right) +  \mathcal{O}(n) \,, 
\end{equation}
Employing the \emph{Master theorem} -- see, for instance, \cite{CLRS_Intro}\footnote{In general, the \emph{Master theorem} states the following: given $a \geq 1$, $b > 1$ and a function $f(n)$, the function $T(n)$ defined on non-negative integers via the recursive relation
	\begin{equation*}
		T(n) = a T\left( \frac{n}{b} \right) + f(n)\,,
	\end{equation*}
	then $T(n)$ asymptotically behaves, for large $n$, as follows:
	\begin{enumerate}
		\item if $0  \leq f(n)  \leq c n^{{\log}_b a - \epsilon}$ for some $\epsilon > 0$, $c > 0$, and for all $n \geq n_0$, then $T(n) \sim n^{\log_b a}$;
		\item if $0 \leq c_1 n^{{\log}_b a } \leq f(n)  \leq c_2 n^{{\log}_b a }$ for some $\epsilon > 0$, some constants $c_1, c_2 > 0$, and for all $n \geq n_0$, then $T(n) \sim n^{\log_b a} \log_2 n$;
		\item if $0 \leq c n^{{\log}_b a + \epsilon} \leq f(n) $ for some $\epsilon > 0$, $c > 0$, and for all $n \geq n_0$, and if $a f\left(\frac{n}{b}\right) \leq C f(n)$ for some $C < 1$, then $T(n) \sim f(n)$.
	\end{enumerate}
} -- one can show that the time complexity scales as $T(n) \sim  \mathcal{O} ( n \log n)$

The worst case occurs when there is no partition at any step of the iteration, with the sole point that gets excluded being the farthest one from the line; then,
\begin{equation}
	\label{eq:QH_Compl_I_b}
	T(n) = T\left( n - 1 \right) + \mathcal{O}(n)\,.
\end{equation}
In this case, the algorithm stops after $n$ iterations, and thus $T(n) \sim \mathcal{O}(n^2)$.

Being the algorithm defined iteratively, the time complexity of the `zero-th step', namely the starting one, with $n = N_s$, gives a measure of the time complexity of the algorithm.
Thus, we see that, \emph{at worst}, the time complexity of the \textsf{WGCHull (I)} subalgorithm is
\begin{equation}
	\label{eq:QH_Compl_I}
	T(N_s) \sim \mathcal{O}(N_s^2)\,.
\end{equation}
This shows that the two-dimensional \textsf{WGCHull (I)} subalgorithm resides in the polynomial time class \textsf{P}.

\subsection{Constructing the convex hull in arbitrary dimensions}
\label{sec:Hull_gend}

\begin{wrapfigure}{r}{0.42\textwidth}
	\centering
	\includegraphics[width=0.4\textwidth]{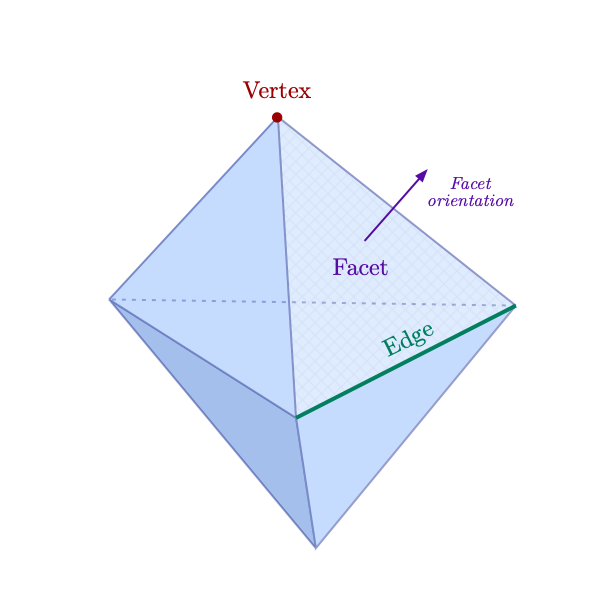}
	\caption{\footnotesize A three-dimensional convex hull.
		\label{Fig:Poly_A}}
\end{wrapfigure}
Before presenting the more general version of the \textsf{Quickhull} algorithm~\cite{BDH}, and to keep this section self-contained, we first recall key definitions and concepts related to convex hulls, following~\cite{ConvPol}.
Consider a set of points $\mathcal{S}$ in $\mathbb{R}^N$ -- in our case, the set given by the charge-to-mass ratio vectors of superextremal particles. The \emph{convex hull} ${\rm conv}( \mathcal{S})$ of the set $\mathcal{S}$ is the intersection of all the convex sets in $\mathbb{R}^N$ that contain $\mathcal{S}$.
In $\mathbb{R}^N$ such convex hulls are polyhedra constructed from the intersection of halfspaces bounded by supporting hyperplanes of the form
\begin{equation}
	\label{eq:Hulld_hyp}
	\mathcal{H}_\alpha: \;{\bf a}^T_{\alpha} {\bf z} + 1 = 0 \qquad \text{with $\alpha = 1, \ldots, m$}.
\end{equation}
These \emph{supporting hyperplanes} are at zero distance from ${\rm conv}( \mathcal{S})$, and define its boundaries.
The points belonging to $\mathcal{S}$ that are not part of the strict interior of ${\rm conv} (\mathcal{S})$ are called \emph{vertices}. 
A \emph{face} $F$ of the convex hull is a subset of ${\rm conv}( \mathcal{S})$ satisfying $F = \mathcal{H}_\alpha \cap {\rm conv}( \mathcal{S})$ for some supporting hyperplane $\mathcal{H}_\alpha$.
An $(N-1)$-dimensional face is called \emph{facet}, while an $(N-2)$-dimensional face is called \emph{edge}, or \emph{ridge}.
For instance, in Figure~\ref{Fig:Poly_A} is a three-dimensional representation of a convex hull.

For the upcoming discussion, we assume that the points in $\mathcal{S}$ are in \emph{general positions}: namely, given any subset of $\mathcal{S}$ containing $N+1$ points, at most $N$ of them lie on the same hyperplane.
Under such an assumption, it can be shown that any convex hull is a simplicial convex. 
Indeed, a \emph{$j$-simplex} is the convex hull of affinely independent $(j+1)$ points; a \emph{simplicial complex} $\mathcal{C}$  is a collection of polytopes in $\mathbb{R}^N$ composed of $j$-simplices as faces, such that every face of any member of $\mathcal{C}$ is also a member of $\mathcal{C}$, and the intersection of any two members is a face of both.

In addition, we adopt the convention that the supporting hyperplanes \eqref{eq:Hulld_hyp} are oriented with outward-pointing unit normals, indicating the direction away from the interior of the hull, as shown in Figure~\ref{Fig:Poly_A}.

A final, key ingredient that we will employ in the following for the inductive construction of convex hulls is the \emph{`beneath-beyond' theorem} \cite{ConvPol}, which we present here in a simplified form. 
Consider a convex hull $C$ in $\mathbb{R}^N$, and a point ${\bf z}$ that does not belong to the convex hull $C$.
Now, let us assume to construct the convex hull ${\rm conv} (C \cup {\bf z})$; a facet $F$ is a facet of the new convex hull ${\rm conv} (C \cup {\bf z})$ if and only if 
\begin{enumerate}[itemsep=0mm]
	\item $F$ was already a facet of the original hull $C$ and the point ${\bf z}$ lies below $F$;
	\item $F$ is newly formed, not being a facet of $C$, and its vertices are ${\bf z}$ and the vertices of an edge of $C$, with an incident facet lying below the point ${\bf z}$ and the other incident facet being above ${\bf z}$.  
\end{enumerate}

\begin{figure}[H]
	\centering
	\includegraphics[width=0.7\textwidth]{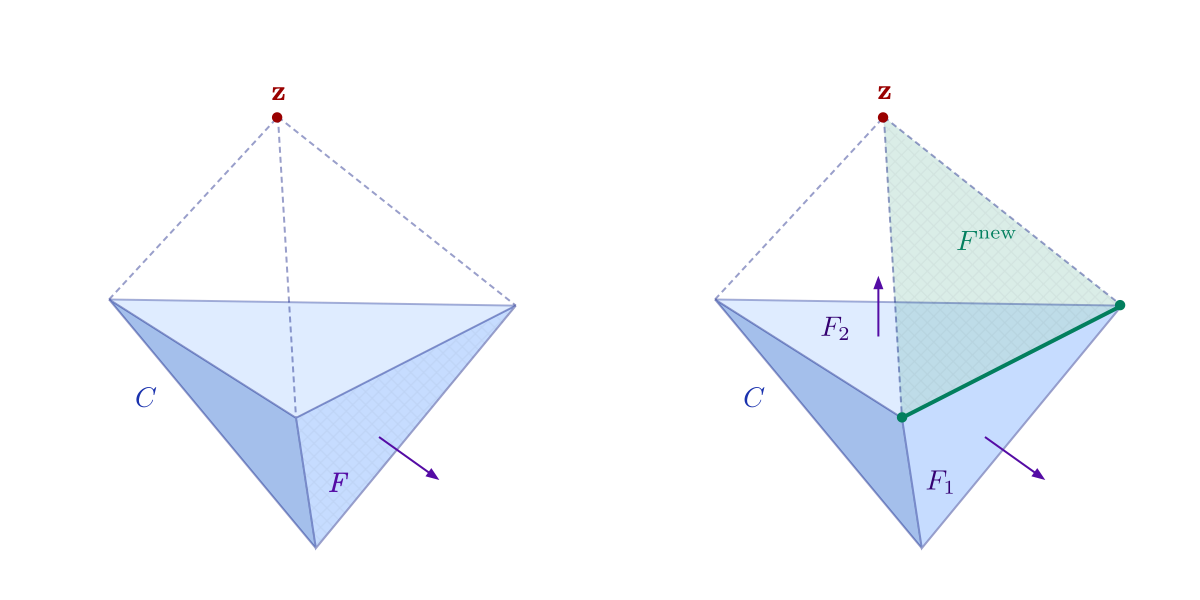}
	\caption{\footnotesize A visual representation of the two cases described by the simplified beneath-beyond theorem.
		\label{Fig:Poly_B}}
\end{figure}

In Figure~\ref{Fig:Poly_B} is a visual, three-dimensional representation of the two cases portrayed by the beneath-beyond theorem. 
The starting convex hull $C$, here a simple tetrahedron, is depicted in light blue, and the external point ${\bf z}$ in red; the resulting convex hull ${\rm conv} (C \cup {\bf z})$ is the one depicted in Figure~\ref{Fig:Poly_A}. 
On the left is an example of a facet $F$ that, while belonging to the original hull $C$, is also part of the final hull ${\rm conv} (C \cup {\bf z})$: the point ${\bf z}$ lies \emph{below} the facet $F$, given its orientation.
On the right is the case of a facet $F^{\rm new}$ that is part of the final hull ${\rm conv} (C \cup {\bf z})$ but does not belong to the original hull.
Indeed, the new facet $F^{\rm new}$ has ${\bf z}$ and two vertices of an edge of $C$, depicted in green, as its vertices; one of the two incident facets to the edge, $F_1$, is \emph{above} ${\bf z}$, while the other, $F_2$, lies \emph{below} ${\bf z}$.

\begin{figure}[t]
	\centering
	\includegraphics[width=\textwidth]{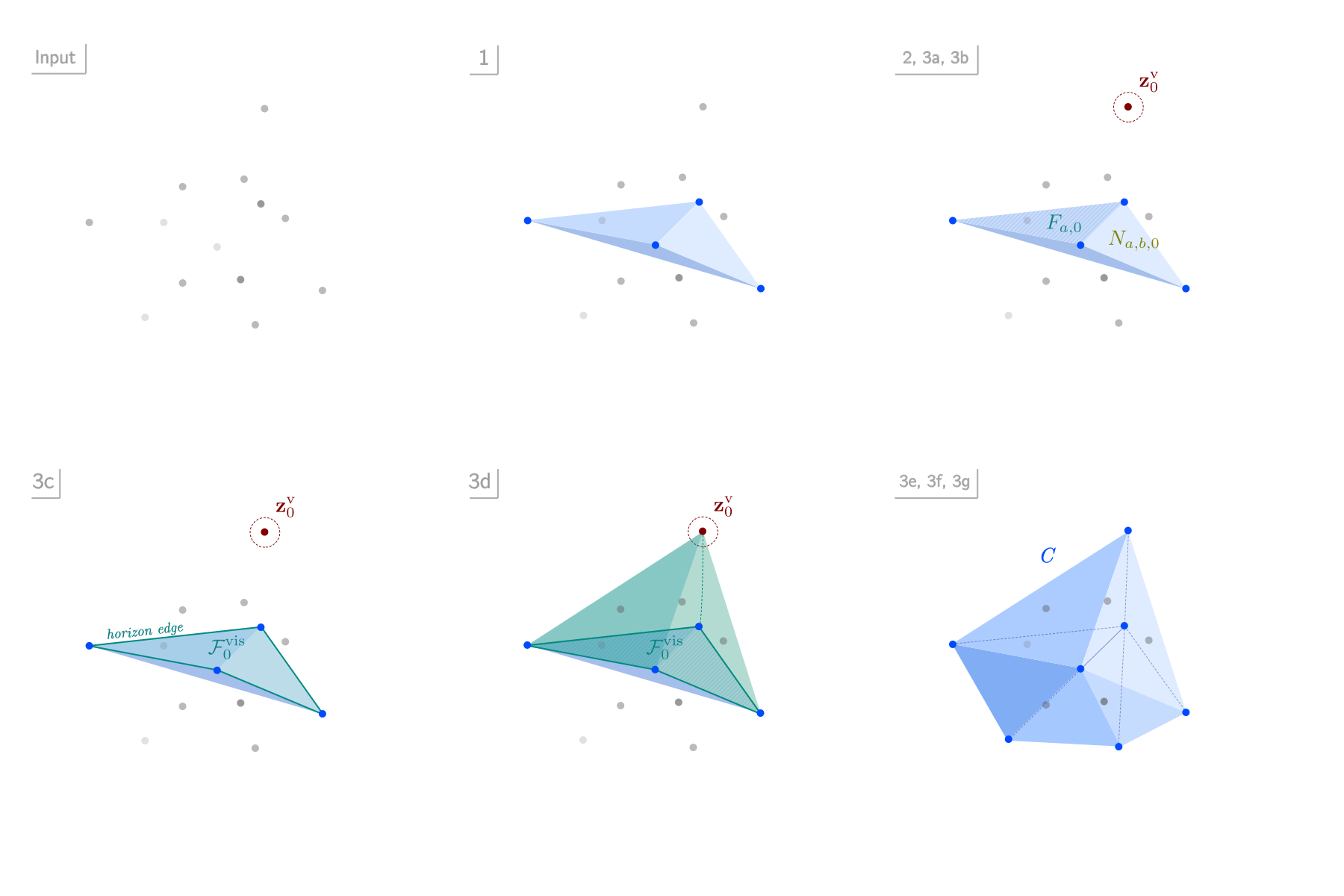}
	\caption{\footnotesize The steps of the generic $N$-dimensional version of the \textsf{Quickhull} algorithm.
		\label{Fig:WGCr_Hulld}}
\end{figure}

With these concepts in place, we now outline the generic $N$-dimensional \textsf{Quickhull} algorithm~\cite{BDH}, a generalization of the algorithm from Section~\ref{sec:WGCr_CH}. 
The input and output remain unchanged; only the steps -- depicted in Figure~\ref{Fig:WGCr_Hulld} -- are adapted for arbitrary $N$:
\begin{enumerate}[itemsep=0mm, wide, labelwidth=!, labelindent=0pt]
	\specialitem{colorloc4}{1} Randomly select $N+1$ charge-to-mass ratio vectors ${\bf z}^{(i)}$ within the set of superextremal particles $\mathcal{S}$, and construct the initial $N$-simplex generated by said vectors. 
	For instance, in three dimensions, this would amount to constructing a tetrahedron.
	The assumptions that the endpoints of ${\bf z}^{(i)}$ are in generic position ensures that the construction of such a simplex is always possible. 
	We call this simplex $S_0$, and we denote $F_{a,0}$, with $a = 1, \ldots, N +1$, its facets. 
	The charge-to-mass ratios ${\bf z}^{(i)}$ lying inside $S_0$ can be neglected, for they are contained in the final hull.
	\specialitem{colorloc4}{2} Consider each of the facets $F_{a,0}$ of the simplex $S_0$ just constructed, and let $\mathcal{H}_{a,0}$ be the hyperplane where the facet $F_{a,0}$ lies. 
	With the help of \eqref{eq:QH_sign_0}, for each of the facets $F_{a,0}$ we identify the charge-to-mass ratios ${\bf z}^{(i)}$ that are \emph{above} the corresponding hyperplane $\mathcal{H}_{a,0}$, and let us call as $\mathcal{A}_{a,0}$ the sets of ${\bf z}^{(i)}$ singled out in this way.
	\specialitem{colorloc4}{3} Let us focus on a single facet $F_{a,0}$ and its set $\mathcal{A}_{a,0}$ of superextremal particles above it:
	\begin{enumerate}[label*=\alph*,itemsep=0mm]
		\specialitem{colorloc7}{3a} Employing \eqref{eq:QH_d_0}, single out the point ${\bf z}_0^{\rm v}$ located at maximum distance from the hyperplane $\mathcal{H}_{a,0}$ where the facet $F_{a,0}$ lies.
		\specialitem{colorloc7}{3b} We want to construct the new hull ${\rm conv}(S_0 \cup {\bf z}_0^{\rm v})$ by using the observations of the beneath-beyond theorem. To this end, initialize a set of facets $\mathcal{F}^{\rm vis}_0$, originally containing only the facet $F_{a,0}$. 
		Then, consider a facet $N_{a,b,0}$, for some index $b$, that is a neighbor of $F_{a,0}$, sharing an edge with it.
		If the facet $N_{a,b,0}$ is \emph{below} the point ${\bf z}_0^{\rm v}$ in the sense of  \eqref{eq:QH_sign_0}, we add $N_{a,b,0}$ to the set $\mathcal{F}^{\rm vis}_0$.
		Thus, loosely speaking, the set of facets $\mathcal{F}^{\rm vis}_0$ collects all the facets of $S_0$ that are `visible' from the point ${\bf z}_0^{\rm v}$. 
		The other facets of $S_0$ that we are disregarding in this step are all above ${\bf z}_0^{\rm v}$ and, by the beneath-beyond theorem, they will be part of ${\rm conv}(S_0 \cup {\bf z}_0^{\rm v})$.
		\specialitem{colorloc7}{3c} Identify the boundary of the set $ \mathcal{F}_{\rm vis}$: this is a collection of edges that we call the `\emph{horizon edges}'.
		\specialitem{colorloc7}{3d} Create new facets (i.e.~$(N-1)$-simplices) having the point ${\bf z}_0^{\rm v}$ as one of their vertices and the opposite edge being one of the horizon edges identified above. We call the new facets so constructed $F_{a,c,1}$, with the index $c$ labeling them.
		\specialitem{colorloc7}{3e} For each of the new facets $F_{a,c,1}$ so obtained, construct their above sets $\mathcal{A}_{a,c,1}$. 
		The points that lie inside ${\rm conv}(S_0 \cup {\bf z}_0^{\rm v})$ can then be neglected since they will be inside the final hull.
		\specialitem{colorloc7}{3f} The facets in $\mathcal{F}^{\rm vis}_0$, which are now internal to the hull, can be eliminated.
		\specialitem{colorloc7}{3g} Repeat the steps from 3.a to 3.f for each of the facets $F_{a,c,1}$, and proceed recursively until the outside sets that would be constructed with step 2 are all empty.
	\end{enumerate}
\end{enumerate}

As with its two-dimensional counterpart in Section~\ref{sec:WGCr_CH}, the general $N$-dimensional \textsf{Quickhull} algorithm outputs the convex hull $C$ formed by the set of superextremal particles $\mathcal{S}$, along with the minimal set of supporting hyperplanes that define its facets. This output can then be passed to the \textsf{WGCHull (II)} subalgorithm from Section~\ref{sec:WGCr_cov} to determine whether the constructed hull fully contains the black hole extremality sphere specified in \eqref{eq:WGC_BHextr_b}.

The computational complexity of this general version of the \textsf{Quickhull} algorithm is less trivial to infer than its two-dimensional counterpart. 
It was already obtained in the original \cite{BDH}, and depends on the input data -- namely, the number of particles $N_s$ in the set $\mathcal{S}$, and the number of independent gauge fields $N$ -- as in \eqref{eq:QH_Compl_WGCHull_I}.

\bibliographystyle{jhep}
\bibliography{references.bib}

\end{document}